\documentclass[preprints,article,accept,moreauthors,pdftex]{Definitions/mdpi} 

\firstpage{1} 
\pubvolume{xx}
\issuenum{1}
\articlenumber{5}
\pubyear{2020}
\copyrightyear{2020}
\history{Received: date; Accepted: date; Published: date}

\usepackage{relsize}

\usepackage{color}

\newcommand{\fr}[1]{\textcolor{black}{#1}}

\Title{Optimal control of colloidal trajectories in inertial microfluidics using the Saffman effect}

\Author{Felix R\"uhle $^{1,}$*\orcidA{}, Christian Schaaf $^{1}$\orcidB{} and Holger Stark $^{1,}$*\orcidC{}}

\AuthorNames{Felix R\"uhle, Christian Schaaf and Holger Stark}

\address[1]{%
$^{1}$ \quad Institut f\"ur Theoretische Physik, Technische Universit\"at Berlin, D-10623 Berlin, Germany 
}

\corres{Correspondence: ruehle@tu-berlin.de (F.R.), holger.stark@tu-berlin.de (H.S.)}

\abstract{
In inertial microfluidics colloidal particles in a Poiseuille flow experience the Segr{\'e}-Silberberg lift force, which drives them
to specific positions in the channel cross section. Due to the Saffman effect an external force applied along the microchannel 
induces a cross-streamline migration to a new equilibrium position. We apply optimal control theory to design the time protocol
of the axial control force in order to steer a single particle as precisely as possible from a channel inlet to an outlet at a chosen target position. We discuss the influence of particle 
radius and channel length and show that optimal steering is cheaper than using a constant control force. Using a single 
optimized control-force protocol, we demonstrate that even a pulse of particles spread along the channel axis can be steered 
to a target and that particles of different radii can be separarted most efficiently.}



\begin{document}

\section{Introduction}

The field of microfluidics is of utmost importance for numerous technological, biochemical, and biomedical applications, especially for inexpensive lab-on-a-chip applications and parallelized or automized studies~\cite{SquiresQuake2005,IsmagilovWhitesides2001,Whitesides2006,WeibelWhitesides2006}.
The experimental realization of high throughput has led to the emergence of inertial microfluidic systems~\cite{DiCarlo2009,AminiDiCarlo2014}, that open up new possibilities. One important characteristic of the
regime of intermediate Reynolds numbers, where flow is still laminar, is the breaking of Stokes reversibility. This leads to self-assembly, such as the famous Segr{\'e}-Silberberg effect discovered by its namesakes in 1961~\cite{SegreSilberberg1961}, where colloids travel to distinct lateral positions in the channel cross section driven by inertial lift forces. Exploiting secondary flow and inertial effects leads to many exciting applications~\cite{DiCarlo2009}, such as enhanced micromixing in curved channels~\cite{SudarsanUgaz2006}, particle separation and filtration~\cite{BhagatPapautsky2008,SajeeshSen2014}, or focusing and self-assembly~\cite{DiCarloToner2009,LeeDiCarlo2010,SchaafStark2019}.
An intriguing aspect is the reaction to external forces pointing along the axial direction of the microchannel. Under creeping flow conditions these forces do not cause lateral migration~\cite{Bretherton1962}, but do so in inertial microfluidic based on the so-called
Saffman effect~\cite{Saffman1965}. Thus, the particles' lateral positions in the cross section of a microchannel can be manipulated with axial external forces, which drive the particles, \textit{e.g.}, via electrophoresis~\cite{KimYoo2009,YuanLi2016,ChoudharyPushpavanam2019}.

Parallel to the experimental progress there have been continuous and fruitful efforts to tackle inertial microchannels via computer simulations~\cite{BazazWarkiani2020}. Here, lift forces and particle dynamics can be probed for different channels and particle types~\cite{ShinSung2011,AsmolovVinogradova2018,SchaafStark2017}, or in complex fluids~\cite{LiArdekani2015,RaoufiWarkiani2019}. Using external forces, optimal and feedback control has been applied to particle separation and steering under inertial microfluidic conditions \cite{ProhmStark2013,ProhmStark2014}, also together with thermal noise. One example is the hysteretic control scheme, as applied by Prohm and Stark~\cite{ProhmStark2014}. Here, particles are periodically forced back to the channel center using the Saffman effect, while the inertial lift force drives them away from the center. Thus, the particle stays within a finite interval around the channel center.

\fr{In this article, we present the theoretical concept to realize
precise particle steering with time-dependent axial control forces. 
The main idea is to steer the particles to different outlets 
of a microchannel in order to achieve particle separation and filtration. We} use the Saffman effect and optimal control theory \cite{IoffeTihomirov1979} 
to design the time protocol 
of the axial control force in order to steer particles from an initial to a target position in a microchannel (see Fig.\ \ref{fig:channel_control}).
These positions are defined, \emph{e.g.}, by inlets and outlets of the microchannel. As an input for the optimization, we employ the lattice-Boltzmann method to simulate particles in Poiseuille flow in order to obtain a whole set of lift-force profiles, depending on the axial control force. Then, we use analytical fit functions to set up a system of ordinary differential equations that yield the particle trajectories. The time-dependent axial control force for optimally steering the particle from an inlet to an outlet 
follows by numerically minimizing a cost functional with respect to the control force under the condition that the target
at the end of the channel is reached. We thoroughly discuss this method of optimal steering and compare it with steering by a constant control force. Using the optimal control-force protocol for a single particle, we demonstrate that even a pulse of particles spread along the channel axis can be steered to a target. Finally, we show how a single optimized control-force protocol can separate
particles of different radii. \fr{Here, we 
go beyond particle separation with constant axial forces suggested in Ref.~\cite{ProhmStark2014}. By using the same time-dependent 
control force for both particles, one can further increase the lateral particle distance at the end of the channel.}

We introduce the theory of inertial microfluidics and the Saffman effect in Sect.~\ref{sec:theory}. 
We describe the setup of our system, the lift force profiles, and the method of optimal control in Sect.~\ref{sec:methods}. 
The results of our study for single and multi-particle steering are presented in Sect.~\ref{sec:results}
and we conclude in Sect.\ \ref{sect.concl}.

\section{Theory - Inertial Microfluidics}
\label{sec:theory}

\begin{figure}
\centering
\includegraphics[width=0.8\textwidth]{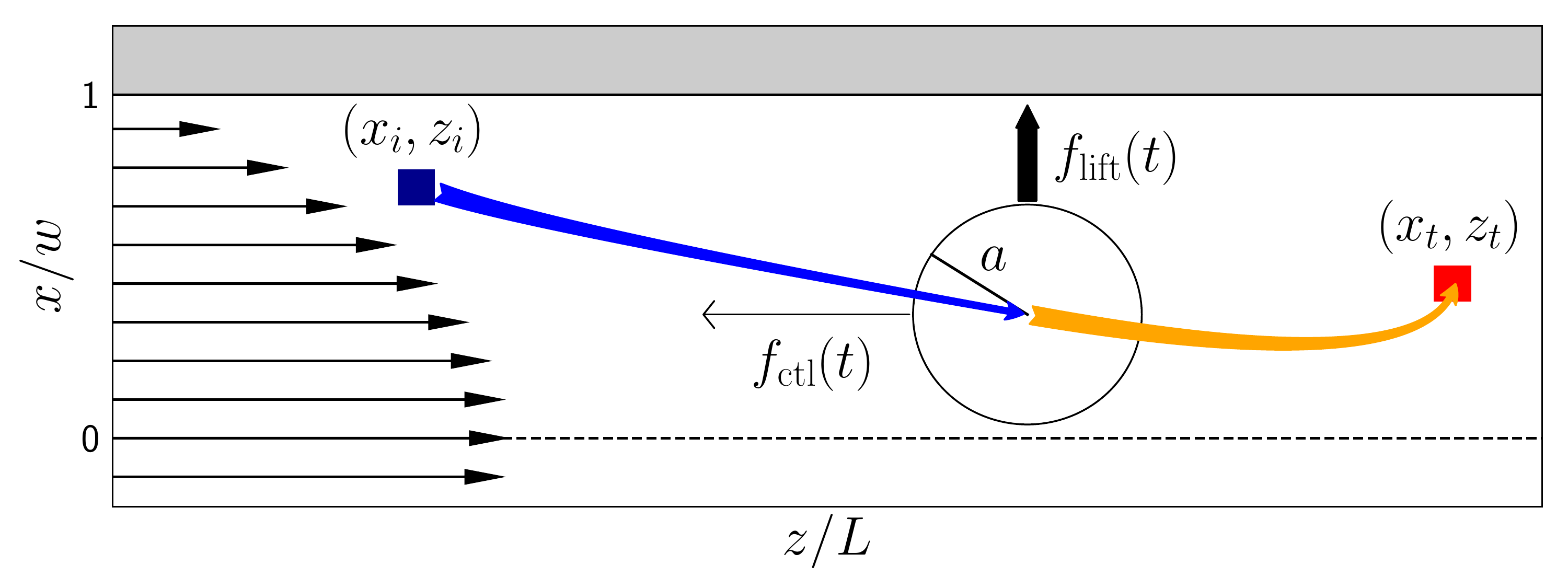}
\caption{Sketch of the model system: A colloid flowing in the $x$-$z$ plane experiences an axial control force $f_{\mathrm{ctl}}(t)$. The occuring Saffman effect changes the lateral lift force $f_{\mathrm{lift}}$ and, thereby, the
colloid can be steered from the initial position $(z_i,x_i)$ to the target $(z_t,x_t)$.
}
\label{fig:channel_control}
\end{figure}

Segr\'e and Silberberg first reported how colloidal particles
self-organize on an annulus under pipe flow conditions~\cite{SegreSilberberg1961} that is located approximately halfway between the channel center and the confining walls. 
Since deterministic lateral motion for rigid particles is impossible under strict creeping flow conditions, this migration results from the inertial term of the Navier-Stokes equation.
Hence, it was termed inertial focusing and rationalized by a so-called inertial lift force \cite{DiCarlo2009,Asmolov1999,HoodRoper2015}.
For channels with rectangular cross sections these equilibrium positions
are either located on the main axes or the diagonals of the cross section, which depends on the particle radius and the cross-sectional aspect ratio (see, for example, Ref.\ \cite{ProhmStark2014}).
If this ratio is sufficiently large, only two stable positions on the short axes exist and
it is sufficient to treat the flowing particle in a two-dimensional plane as sketched in Fig.\ \ref{fig:channel_control}.

The Poiseuille-flow profile in a rectangular channel is known analytically \cite{Bruus2008}. Since the flow field along the channel 
axis obeys $\mathbf{u} = u(x,y)\mathbf{e}_z$, the convective term of the Navier-Stokes equations vanishes, and for the stationary 
case the Stokes equations are \fr{recovered}. Restricting fluid flow in the cross section to $x\in(-w,w)$, $y\in(-h,h)$ with $h>w$ and using 
no-slip boundary conditions at the channel walls, $u(x=\pm w,y) = u(x,y=\pm h) = 0$, one can write the solution as a Fourier series expansion~\cite{Bruus2008}
\begin{equation}
\mathbf{u}(x,y) = \dfrac{16w^2 \Delta p}{\pi^3\eta L}\mathlarger{\mathlarger{\sum_{n=0}^{\infty}}}(-1)^n \dfrac{1}{(2n+1)^3}\left[1-\dfrac{\cosh\left(\frac{(2n+1)\pi}{2w}y\right)}{\cosh\left(\frac{(2n+1)\pi}{2w}h\right)}\right]\cos\left(\frac{(2n+1)\pi}{2w}x\right).
\end{equation}
Here, a constant pressure gradient $\Delta p/L$ is used and the dynamic viscosity of the fluid $\eta$. When we employ this analytical formula in our numeric calculations, we truncate the series after $n=100$. The maximum flow velocitiy $U_m$ is reached at the center $(x,y)=(0,0)$ of the channel. It is determined by the choice of the Reynolds number
$\mathrm{Re} = \rho U_m 2w / \eta$, where $\rho$ is the fluid density and $2w$ the width of the channel.

Inertial effects become observable if a colloid is subjected to a Poiseuille flow at finite Reynolds numbers $\mathrm{Re}$. 
This initiates a lift force acting on the colloid, which can be controlled via the Saffmann effect by applying an additional axial 
control force (see Fig.\ \ref{fig:channel_control}).
We introduce those in the following.

\subsection{Lift force}
Since the discovery of inertial focussing different scaling laws for the dependence of the inertial lift force on particle radius $a$ and Reynolds number have been derived
\cite{HoLeal1974,SchonbergHinch1989,Asmolov1999,DiCarloToner2009,HoodRoper2015}. 
For example, Ho and Leal calculated the lift force for small particle radius ($a \ll w$) and small 
particle Reynolds number $\mathrm{Re}(a/w)^2$.
They arrived at the scaling law $f_\mathrm{lift}\sim\mathrm{Re}^2(a/w)^4$
\cite{HoLeal1974}, whereas numerical simulations at finite particle sizes arrived at $f_\mathrm{lift}\sim(a/w)^3$ in the channel center for particle sizes $a<w$~\cite{DiCarlo2009}. To correct for finite particle size and Reynolds number, often the lift coefficient $f(a,Re)$ is introduced~\cite{DiCarlo2009}. 
In particular, it has been observed that the scaling exponent for the lift force as a function of the particle radius depends on the lateral position in the channel~\cite{DiCarloToner2009,DiCarlo2009}. 
Importantly, the fixed points of the lift-force profiles indicating stable equilibrium positions change considerably with the geometry of the channel cross section (see, for example, Ref. \cite{ProhmStark2014}).
In the following, we calculate the lift-force profiles numerically using lattice-Boltmann simulations as shortly explained in Sec.\ \ref{subsec.LB}.
Typical examples for a zero axial control force are presented in Fig.\ \ref{fig:lift_force_profiles}, left.

The net inertial lift force is often described as a balance of two contributions.
They arise from a stresslet that is the leading force distribution on the particle surface under a shear-flow gradient \cite{HoLeal1974,AminiDiCarlo2014}: 
the reflection of the stresslet from the channel wall induces a force, which pushes the particle away from the wall,  whereas the interaction with the shear gradient transports particles to regions of larger shear, which is towards the wall in case of a Poiseuille flow. 

\subsection{Saffman effect}
\label{sec:saffman}
\begin{figure}
\centering
\includegraphics[width=0.49\textwidth]{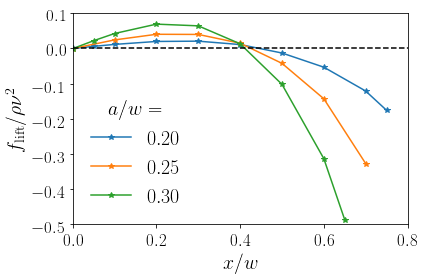}
\includegraphics[width=0.49\textwidth]{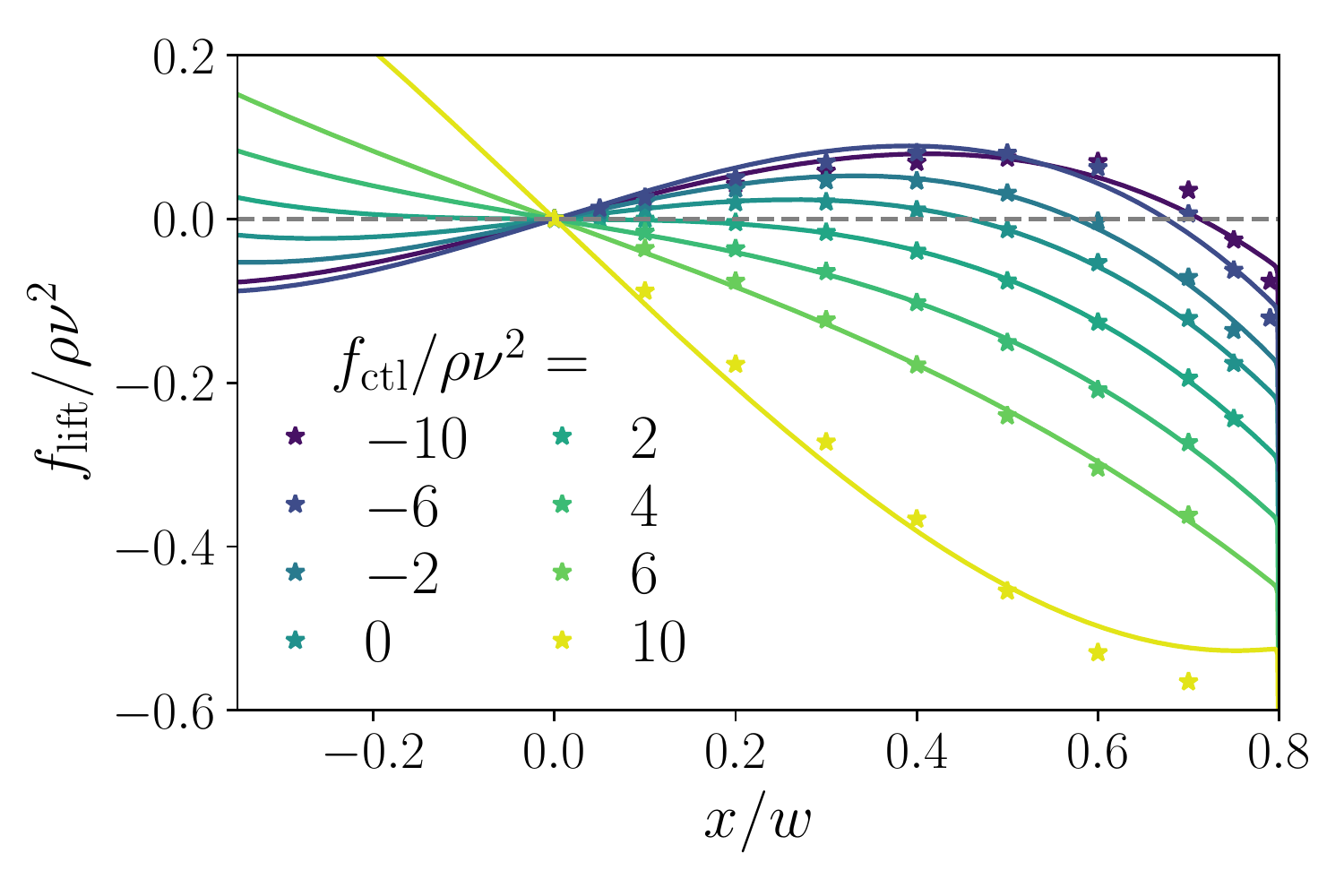}
\caption{Left: Lift-force profiles along the positive $x$ axis for different 
particle radii $a/w$ at zero axial control force, $f_{\mathrm{ctl}} = 0$.
Note the larger strength of the lift forces for larger colloids and the shift of the stable fixed point. The force unit $\rho \nu^2$ uses fluid density $\rho$  and kinematic viscosity 
$\nu = \eta / \rho$.
Right: Lift-force profiles for a colloid with radius $a=0.2w$ at different axial control forces and least-square fits using 
eqs.\ (\ref{eq:fit}) (solid lines). 
In both cases the Reynolds number $\mathrm{Re} = 10$ is used.
}
\label{fig:lift_force_profiles}
\end{figure}

Applying an additional axial control force to the colloidal particle speeds up or slows it down relative to the local Poiseuille flow velocity.
This modifies the slip velocity field close to the particle surface and at finite Reynolds numbers creates an  additional \textit{lateral} contribution to the lift force, described by Saffman~\cite{Saffman1965}. It depends on the shear rate $\gamma$, rather than the shear gradient, and 
was calculated to be $f_S\sim v a^2 \gamma^{1/2}$ in bulk at small Reynolds numbers, where $v$ is the difference between local flow field and particle velocity.
Figure\ \ref{fig:lift_force_profiles}, right demonstrates how the lift-force profile changes, when a control force is applied.
The stable fixed point ($f_\mathrm{lift}=0$ with a negative slope) moves from the zero-force position ($f_{\mathrm{ctl}} = 0$) either to the wall or to the channel center depending on whether the control force is applied along the flow direction ($f_{\mathrm{ctl}} < 0$) or against it
($f_{\mathrm{ctl}} > 0$), respectively.
The respective stable equilibrium positions are plotted in Fig.~\ref{fig:stable_fixed_points}, left, also for different particle radii.
Because the inertial lift force
grows more strongly with the radius than the Saffman 
force, higher axial control forces are necessary to move the fixed point for larger particles towards the center.
Consequently, the curves in Fig.~\ref{fig:stable_fixed_points}, left become flatter for larger particle radii.

\section{Methods}
\label{sec:methods}

\subsection{Setup}
\label{subsec.setup}

We consider a rectangular microchannel with Poiseuille flow at Reynolds number $\mathrm{Re}=10$. The channel 
has an aspect ratio $w:h$ of 1:2, where $x\in(-w,w)$ and  $y\in(-h,h)$. The length of the channel is $L$
with $z\in(0,L)$, which we vary in the following. At 
such an intermediate Reynolds number, the regime of inertial microfluidics is reached 
and solid colloids in a Poiseuille flow self-organize towards distinct lateral focus positions~\cite{Saffman1965}. 
For the aspect ratio chosen here, they equilibrate to the plane $y=0$ and, therefore, we only consider the particle dynamics in the $x$-$z$-plane \cite{AminiDiCarlo2014,BhagatPapautsky2008,GossettDiCarlo2012Small}. \fr{Indeed in our lattice-Boltzmann simulations the position $y=0$ is stable, i.e., a particle is immediately driven back once it leaves the center plane. The stable fixed point at this location was numerically determined in Ref.~\cite{ProhmStark2014}.  
Switching on the axial force, the induced Saffman force is strongest along the $x$ axis since in this direction the
velocity gradients are largest. Therefore, we expect the particle to stay in the center plane.}
The stable equilibrium positions on the $x$ axis
($f_\mathrm{lift} = 0$)
depend on particle size 
as we show in Fig.~\ref{fig:lift_force_profiles}, left. 
There is also a dependence on $\mathrm{Re}^2$~\cite{SchaafStark2019},
which we do not further explore here.
These positions are reached after the colloid has been advected for a sufficiently large axial distance $L_f$ without any external forcing. 
Di Carlo and co-workers mention an estimate for this length
\cite{DiCarlo2009,AminiDiCarlo2014},
\begin{equation}
L_f = \dfrac{\pi\nu w^2}{f_L U_m a^2},
\label{eq:focus_length}
\end{equation}
with the maximum flow velocity $U_m=\nu\mathrm{Re}/(2w)$, kinematic fluid viscosity $\nu$, and particle radius $a$. 
For our aspect ratio $w/h=0.5$, Ref.\ \cite{DiCarlo2009} gives a  lift coefficient $f_L = 0.05$. Furthermore, using $\mathrm{Re} = 10$, and $a/w=0.2$, in eq.\ (\ref{eq:focus_length}), we obtain the focus length
$L_f\approx 314 w$. 
Now, applying an additional lateral force along the $x$ direction, one can optimally steer particles to any position on the 
$x$ axis as we showed in Ref.~\cite{ProhmStark2013}.

\begin{figure}
\centering
\includegraphics[width=0.49\textwidth]{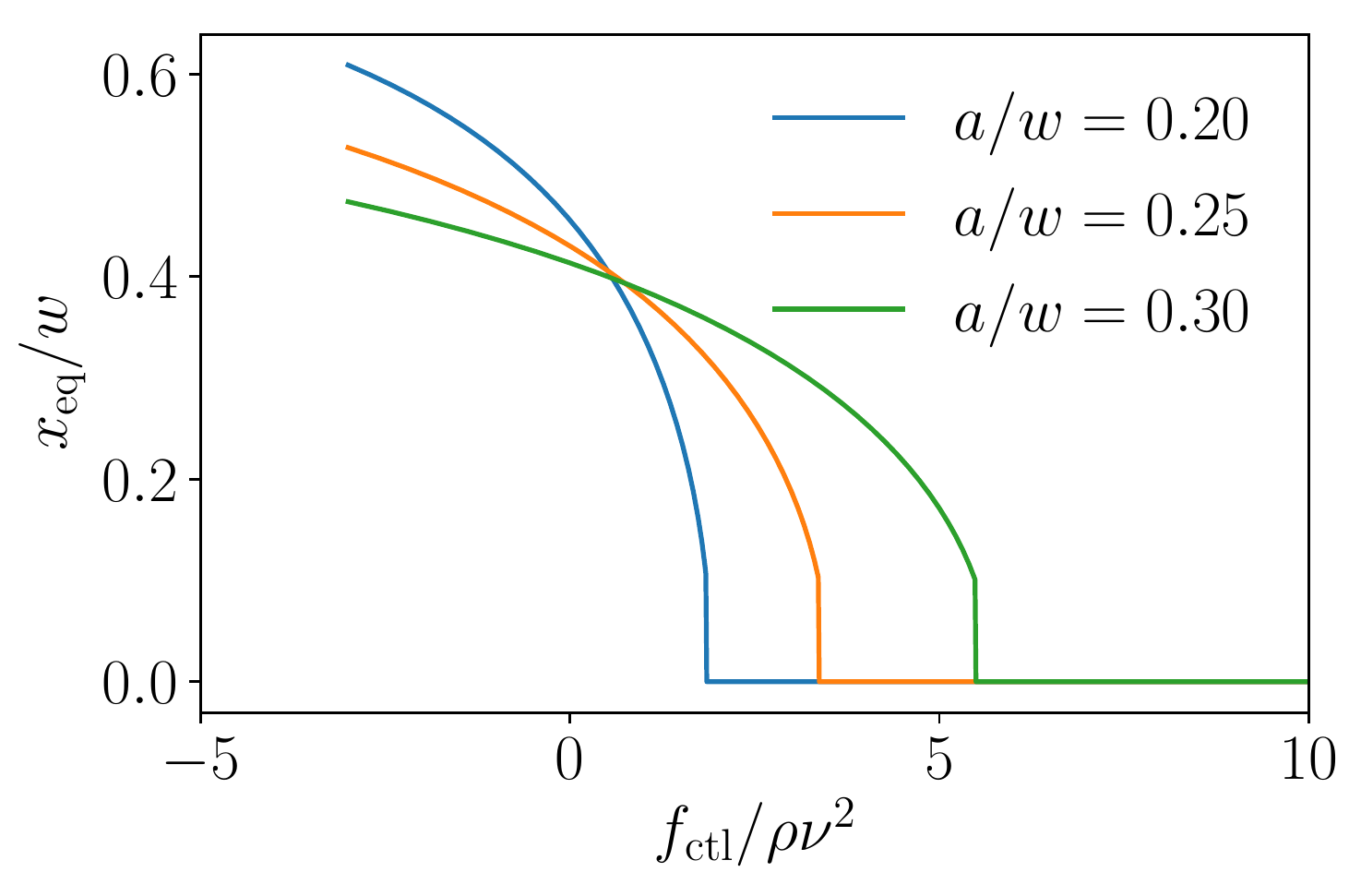}
\includegraphics[width=0.49\textwidth]{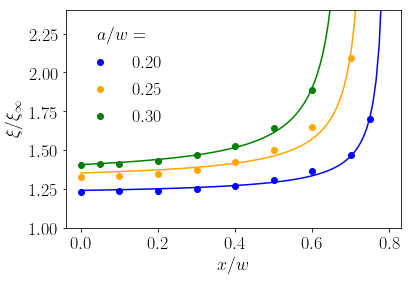}
\caption{Left: Stable equilibrium positions (fixed points) as a function of control force for different 
particle radii $a/w$. They were determined 
using the analytical fits 
to the relevant lift-force profiles in Fig.~\ref{fig:lift_force_profiles}, right. Right: 
Hydrodynamic friction coefficients
relative to the bulk value $\xi_{\infty} = 6\pi \eta a$ plotted versus the lateral particle position for different particle radii. The solid lines are fits using eq.\ (\ref{eq:friction}).
}
\label{fig:stable_fixed_points}
\end{figure}

Here, we propose an alternative strategy 
for optimal steering using the Saffman effect. We apply an axial control force and thereby 
modify the lift-force profile as demonstrated in Fig.~\ref{fig:lift_force_profiles}, right. 
In Fig.~\ref{fig:stable_fixed_points} we show how the stable equilibrium position now depends on the control force.
Then, the idea is to use a time varying axial control force, which can be realized, for example, by  electromagnetic 
fields~\cite{YuanLi2016}, for optimal steering. \fr{The goal is} to optimally steer a particle from an inlet, which is located at the start position $(z_i,x_i)$, towards a target $(z_t,x_t)$ fulfilling a criterion of optimality as we will outline below. 
To implement this approach, we first need lift-force profiles $f_\mathrm{lift}(x,f_\mathrm{ctl})$ for different control forces as well as friction coefficients $\xi(x)$ for different particle sizes, which we determined with the help of lattice-Boltzmann simulations,
and approximate them with appropriate fit functions (Sect.\ \ref{subsec.LB}). They are then used in dynamical equations
for the particle motion, which we solve with explicit Euler integration in order to determine the optimal steering path 
(Sect.\ \ref{sec:system_control}).

\subsection{Profiles for lift forces and friction coefficients}
\label{subsec.LB}

Our lattice-Boltzmann simulations (including the immersed-boundary method) \cite{ChunLadd2006,DuenwegLadd2009} of single 
colloids in a microchannel in the inertial regime are described in detail in Refs.\ \cite{ProhmStark2014,SchaafStark2019}, where 
we also explain how to determine inertial lift forces for each particle position.

The simulated lift-force profiles for different particle radii and channel Reynolds number $\mathrm{Re} = 10$ are shown in Fig.~\ref{fig:lift_force_profiles}, left.
They display the well-known behaviour of inertial focusing: Colloids are driven away from the unstable fixed point at the origin and towards their 
stable equilibrium positions (stable fixed points) between the channel center and the wall, 
which depend on the particle radius. As in Ref.~\cite{ProhmStark2013}
we perform a least-square fit of the lift-force profiles to a third-order polynomial of odd degree together with a wall-repulsion term as the particle approaches the walls.
Additionally, we now also apply this fit to the dependence of the lift force on the axial control force $f_\mathrm{ctl}$ using  coeffients that are
second-order polynomials in $f_\mathrm{ctl}$. Thus, the functional form for the fit of our lift-force 
profiles is as follows
\begin{eqnarray}
f_\mathrm{lift}(x,f_\mathrm{ctl}) & = & \phi_1(f_\mathrm{ctl})x + \phi_3(f_\mathrm{ctl})x^3 + \phi_w f_w(x) \nonumber \\
\phi_1(f_\mathrm{ctl}) & = & a_1 f_\mathrm{ctl}^2 + b_1 f_\mathrm{ctl} + c_1 \nonumber \\
\phi_3(f_\mathrm{ctl}) & = & a_3 f_\mathrm{ctl}^2 + b_3 f_\mathrm{ctl} + c_3 \\
f_w(x) & = & \dfrac{1}{x-(1+\delta)x_w} +  \dfrac{1}{x + (1+\delta)x_w}, \nonumber
\label{eq:fit}
\end{eqnarray}
where 
$x_w = w-a$ and we use $\delta = 10^{-3}$ for numerical stability. 
As Fig.~\ref{fig:lift_force_profiles} demonstrates, the fit function works well, in particular, for non-zero control force, as long as the region immediately at the wall is avoided.
Note that we did not attempt to include the dependence on particle radius in our fit function,
but instead perform a separate fit for each particle size. We do this in order to limit the number of parameters.

We also determined the friction coefficients of the particles in the lattice-Boltzmann simulations and plot their values 
as a function of the lateral position in Fig.\ \ref{fig:stable_fixed_points}, right for three different colloidal radii. The presence of the 
walls is clearly visible.
We fit the position-dependent friction coefficient $\xi$ by the function
\begin{equation}
\xi(x) = \xi_\infty \Big(d_1 + d_2 \dfrac{a}{(w - a)- \vert x\vert}\Big),
\label{eq:friction}
\end{equation}
where $\xi_{\infty} = 6 \pi \eta a$ is the bulk friction coeffcient and $d_1$, $d_2$ are fit parameters. The fits as solid
lines are presented in Fig.~\ref{fig:stable_fixed_points}, right.

\subsection{Dynamical system and optimal control}
\label{sec:system_control}

Using the axial control force $f_\mathrm{ctl}$, the fitted lateral lift force $f_\mathrm{lift}$, and the friction coefficient $\xi$,
the overdamped motion of the steered particle in the Poiseuille flow profile $u(x) = u(x,y=0)$ is governed by the following 
differential equations disregarding any thermal noise:
\begin{eqnarray}
\label{eq:diff}
\dot{z} & = & u(x) + \dfrac{1}{\xi(x)}f_\mathrm{ctl}(t) \\
\dot{x} & = & \dfrac{1}{\xi(x)}f_\mathrm{lift}\left(x,f_\mathrm{ctl}(t)\right).
\end{eqnarray}
Here, $z$ is the coordinate \fr{along} the channel and $x$ in lateral direction. As described above, the size of the suspended particle influences the lateral motion implicitly via the friction coefficient and the fitted function for the lift force. In axial direction we 
do not consider that a force-free colloid is slower than the streaming fluid but simply set this velocity to the 
Poiseuille flow velocity $u(x)$. We note that the dynamics of the colloid is always confined to one half of the channel. At $x=0$ the lift force - and hence the total lateral force - is exactly zero, therefore it is impossible for the colloid to cross the center line.

Solving these equations for a given time protocol $f_\mathrm{ctl}(t)$ of the axial control force,
determines $x(t)$ and $z(t)$. We are looking for an optimal protocol $f_\mathrm{ctl}^*(t)$, which steers a particle as close as possible to the target $(z_t,x_t)$ at end time $T^*$. 
Thus, we define the cost functional~\cite{IoffeTihomirov1979}
\begin{equation}
\label{eq:cost_functional}
J[f_\mathrm{ctl}(t),T] = \dfrac{c_x}{2}\Big\vert x_t-x(T)\Big\vert^2 + \dfrac{c_z}{2}\Big\vert z_t - z(T)\Big\vert^2 + \dfrac{\varepsilon}{2}\int_{t_0}^{T} |f_\mathrm{ctl}(t)|^2dt,
\end{equation}
and obtain the optimal steering control force 
by minimizing the cost functional with respect to $f_\mathrm{ctl}(t)$:
\begin{equation}
f_\mathrm{ctl}^*(t) = \underset{f_\mathrm{ctl}}{\mathrm{arg\,min}} J \, .
\end{equation} 
For the total duration $T$ of the trajectory, which is 
undetermined on the right-hand side of eq.~\eqref{eq:cost_functional}, the algorithm also finds an optimum $T^*$. 
In concreto, we set $T^* = N \Delta t^*$, where  $\Delta t^*$ is the time step of our time discretization, choose a constant 
$N$, and determine $\Delta t^*$ together with the force protocol by minimizing the cost functional. We always choose a control force that is constant in time as our initial function.
When optimizing the single-particle trajectories, we choose $N=500$;
while when looking at a pulse of colloids in Sect.~\ref{sec.pulse}, we take $N=2500$ since a higher resolution in axial direction is required.
The functional in eq.\ \eqref{eq:cost_functional} includes a regularization term, where we integrate over the square of the total force because otherwise arbitrarily large forces would be permissible. Keeping 
this regularization term low, decreases the energy cost for steering the particle along a specific trajectory, for example, electrophoretically by applying an electric field \cite{YuanLi2016}.
We weigh the cost of deviating from the target area 
differently for the $x$ and $z$ coordinates, because the velocities differ strongly. The control parameters $c_x,c_z$ and 
$\varepsilon$ have to be adapted manually to find the right balance between the cost of higher forces versus the precision of steering. When optimizing the single-particle trajectories, we choose $c_x=14000$, $c_z=0.3$, and $\varepsilon=0.003$. For solving the differential equations, we use explicit Euler integration. With this we optimize the discretized cost functional $J[f_\mathrm{ctl}(t),T]$ using a sequential quadratic programming (sqp) algorithm~\cite{NocedalWright2006Chap18} provided by the 
package \textit{fmincon} from \textit{MathWorks'} software matlab (Release R2019b)~\footnote{\fr{https://mathworks.com/help/optim/ug/constrained-nonlinear-optimization-algorithms.html}}. \fr{This robust and efficient method uses a Lagrangian representation of a constrained problem: $\mathcal{L}=f(x) + \lambda^T c$, where $x$ is the vector of unknowns, $c$ the vector of 
equality constraints and $\lambda$ the vector of Lagrange multipliers. For the optimal solution $x^*$, the gradient of $\mathcal{L}$ vanishes, which yields a nonlinear equation. Here, $x^*$ can be approximated by iteration,
$x_{i+1} = x_i + d_i$, where the differences $d_i$ are determined in each step by a simpler approximated quadratic problem, for which standard quadratic solvers are available~\cite{NocedalWright2006Chap18}. We do not calculate the necessary gradients to the functional ourselves but leave this to the software matlab.}

We also seek to maximize the lateral distance between two colloids,  when approaching the target 
with axial coordinate $z_t$.
Therefore, in the cost functional of eq.~\eqref{eq:cost_functional} we use $\dfrac{c_x}{2}\big\vert \Delta x(T) - \Delta x_f \big\vert^2$ for the first summand, where $\Delta x(t)$ is the lateral distance between two colloids at time $t$ and $\Delta x_f$ is the distance aimed for. 
In the second term involving the $z$ coordinate we add up the contributions from the two particles.
The coefficient $c_z$ takes the same values for both colloids. For the two-particle optimization we use $c_x=500$, $c_z=30$, 
and $\varepsilon=0.01$.

We always obtained smooth solutions when solving the unconstrained problem.
Due to the diverging repulsive lift force close to the walls, it was not necessary to use a numerical constraint for the
lateral coordinate $x$ in order to keep the particle within the channel. The sqp algoritm required $75$ iterations on average in order to converge for the single-particle steering. For the separation of two particles, $161$ iterations were required on average.

\section{Results and Discussion}
\label{sec:results}

In the following we investigate steering strategies for single and multiple particles that make use of the aforementioned inertial
lift forces. First, we discuss steering with a constant axial control force
and then the outcome of our optimal control
scheme. Using the results from single-particle steering, we investigate the implications for a particle pulse spread along the channel axis.
Finally, we use optimal control to find control forces that maximize the separation of two particles  so that they can be carried off at different outlets of a channel.

\subsection{Steering with constant axial control forces}
\label{sec:constant_force}
Exploiting the Saffman effect, the easiest approach to steer a particle to a target position is to use constant axial control forces. 
They shift the stable fixed point of the lift-force profile and thereby, in principle, the equilibrium position of a colloidal particle can be adjusted arbitrarily, as shown in Fig.\ \ref{fig:stable_fixed_points}.
In the same way, also colloids of different sizes can be well separated within the microfluidic channel using a constant control force \cite{ProhmStark2013}.
As Fig.\ \ref{fig:stable_fixed_points} shows this is achieved by choosing the control force such that the smaller particle (\emph{e.g.}, $a=0.2w$) is pushed to the center while the larger particle still keeps a noticeable distance from the center.

\begin{figure}
\centering
\includegraphics[width=0.49\textwidth]{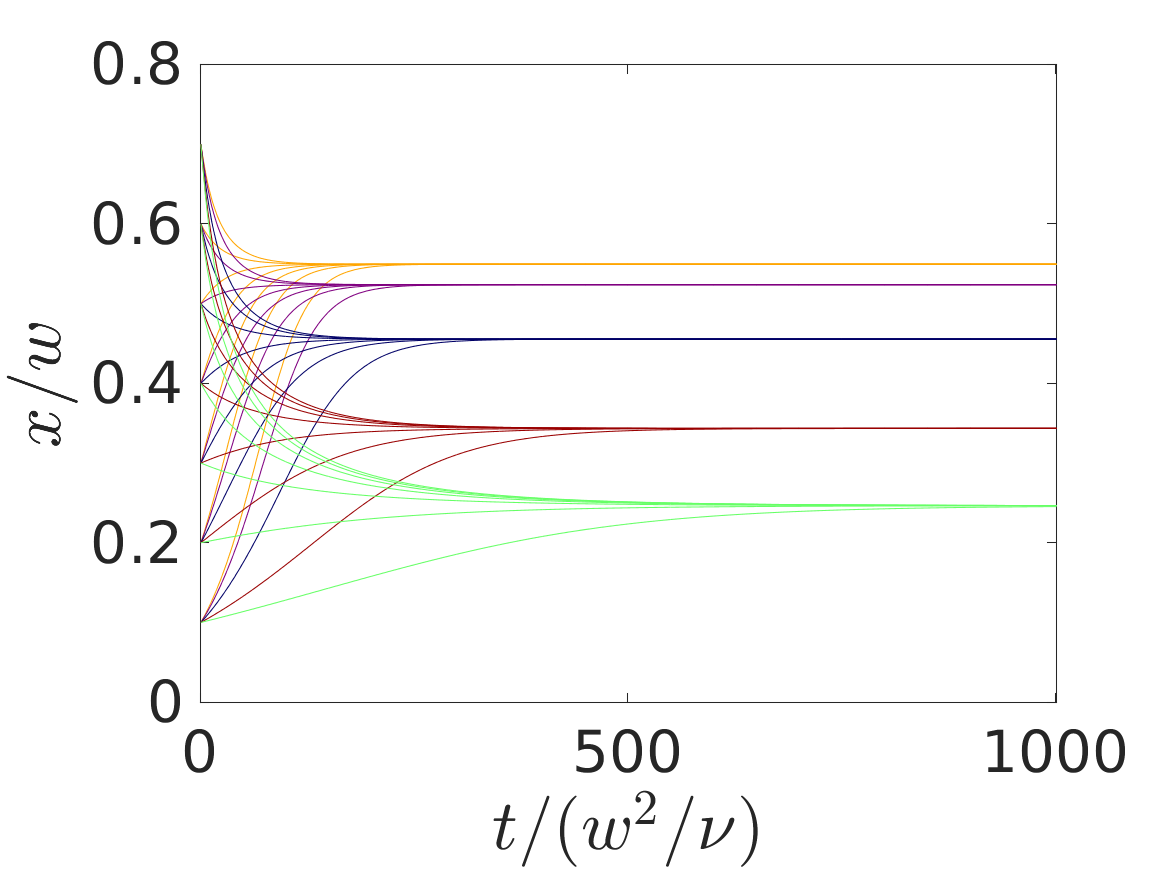}
\includegraphics[width=0.49\textwidth]{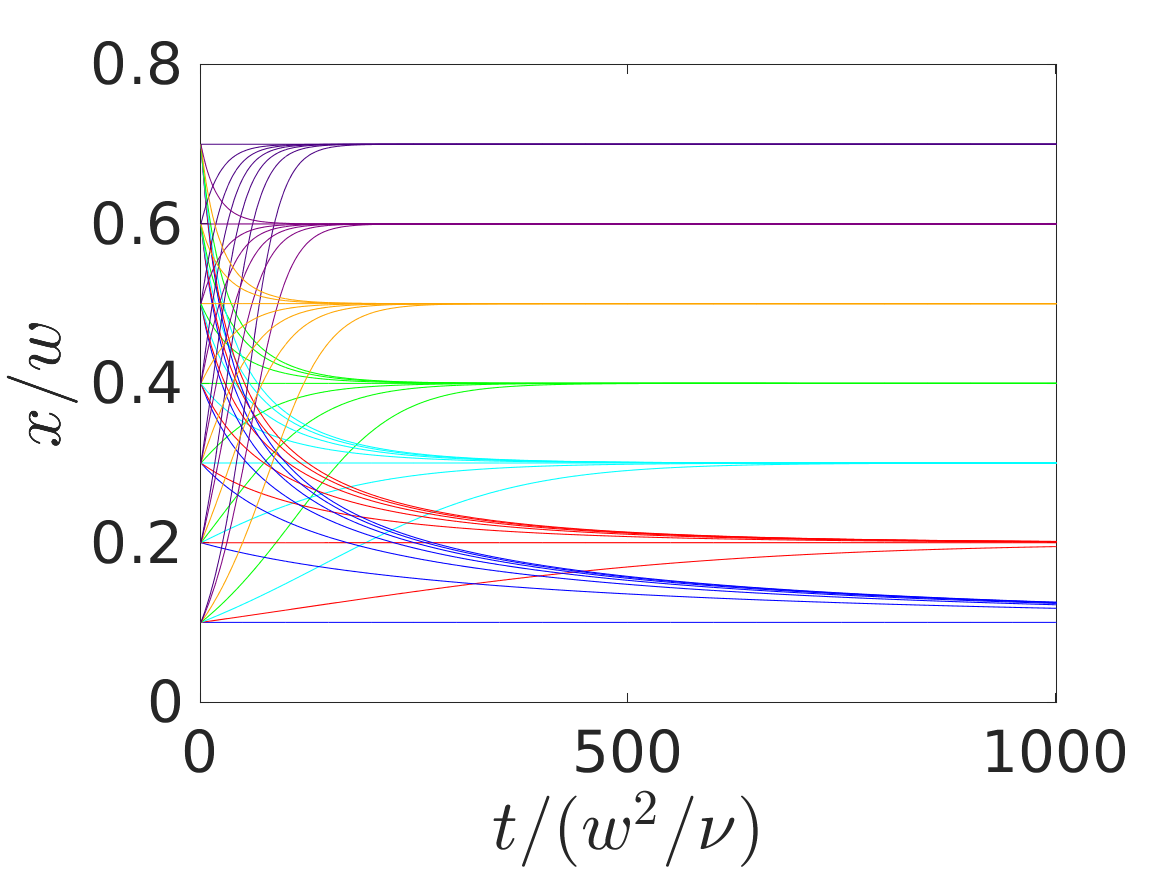}
\caption{Inertial focusing of a colloid with radius $a=0.2w$ starting from different initial positions and using constant control forces. Lateral position versus 
time is plotted. Left: for constant control forces $f_\mathrm{ctl}/(\rho\nu^2) = -1.5,-1.0,0,1.0,1.5$ (which give an increasing 
equilibrium $x$ position or stable fixed point). Right: for adjusted constant control forces such that the stable fixed point 
assumes the values $x/w=0.1,0.2,...,0.7$. 
}
\label{fig:relaxation}
\end{figure}   

In Fig.\ \ref{fig:relaxation} the lateral positions for the moving colloid are plotted versus 
time for specific control forces (left) and when the forces are adjusted to give specific final positions (right). Negative forces point along the 
direction of the channel flow and thus drive the equilibrium position closer to the wall, whereas positive forces slow 
down the colloid and thereby induce motion towards the channel center. 
A closer inspection of the two plots shows that the necessary travel time for focusing varies with the initial position.
However, more pronounced is the dependence on the final position as Fig.~\ref{fig:relaxation}, right demonstrates. 
Fixed points closer to the channel center are reached later than those closer to the wall. 
This is consistent with the fact that the Saffman effect (or the shear-induced lift force) increases with the shear rate, which is larger close to the walls. Thus, using the Saffman effect for steering requires less time 
the closer the final position is situated to the wall.
In Sect.\ \ref{subsec.setup} we evaluated the focus length $L_f \approx 314 w$ for our setup. Indeed, it gives a good estimate 
for the focus length in our simulations at zero control force. When plotting the lateral positions of Fig.\ \ref{fig:relaxation} versus 
the traveled axial distance $z$ instead of time, the curves look very similar to the ones in Fig.\ \ref{fig:relaxation}, even though
particles closer to the center (small $x$) should flow faster and the curves should stretch even farther close to the centerline.
However, to reach these targets, the control force has to act against the flow and therefore decreases the axial flow velocity
of the particles.

\fr{In conclusion, in order to relax to the adjusted  equilibrium position with the constant-force strategy, considerably
longer travel times and axial distances are required for targets that lie close to the channel center. Therefore, long enough channels need to be used in order for the strategy to work. Thus,} it is not possible to use a single channel length to steer particles to different lateral target positions using the constant-force 
strategy. Furthermore, steering with a constant axial force means, it has to be maintained for the whole trajectory. 
We compare the cost of this constant-force scheme with the optimal control scheme in the next section, which also allows to operate with channels of one length.

\subsection{Optimal control of single colloids}
\label{sec:control_single}

We now turn to the optimization problem for the cost functional $J[f_\mathrm{ctl}(t),T]$ of eq.\ (\ref{eq:cost_functional})
set up in Sec.\ \ref{sec:system_control}. This will provide us with a time-dependent control force and the particle trajectory in the
$x$-$z$ plane. In the following, we provide optimal solutions of the cost functional for different start and end positions in the channel. 
The optimization procedure is applied to two axial target positions at $z_t = 300w$ and $z_t = 500w$.
Furthermore, three particle sizes with radii $a=0.2w$, $0.25w$, and $0.3w$ are considered.
Since the initial position is always set at $z_i = 0$, we call $z_t$ channel length for short.

\begin{figure}
\centering
\includegraphics[width=0.40 \textwidth]{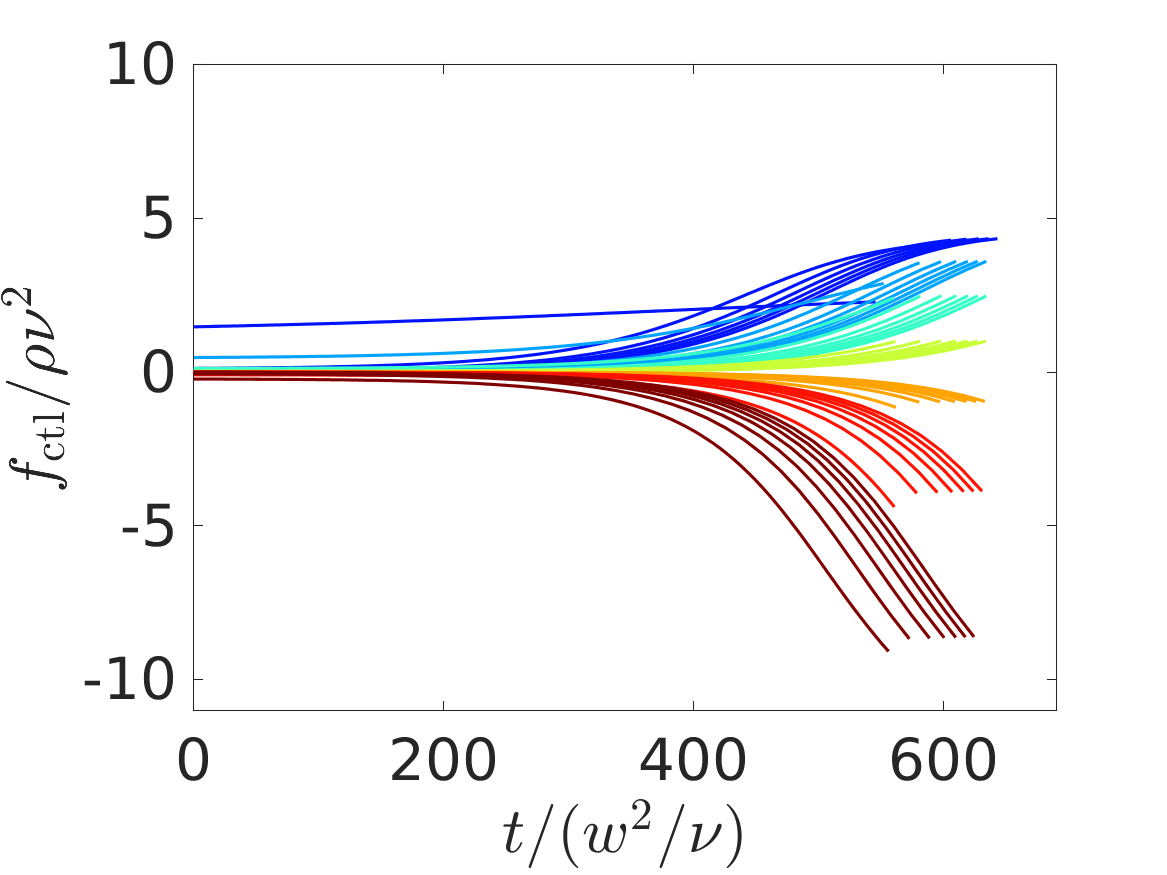}
\includegraphics[width=0.40 \textwidth]{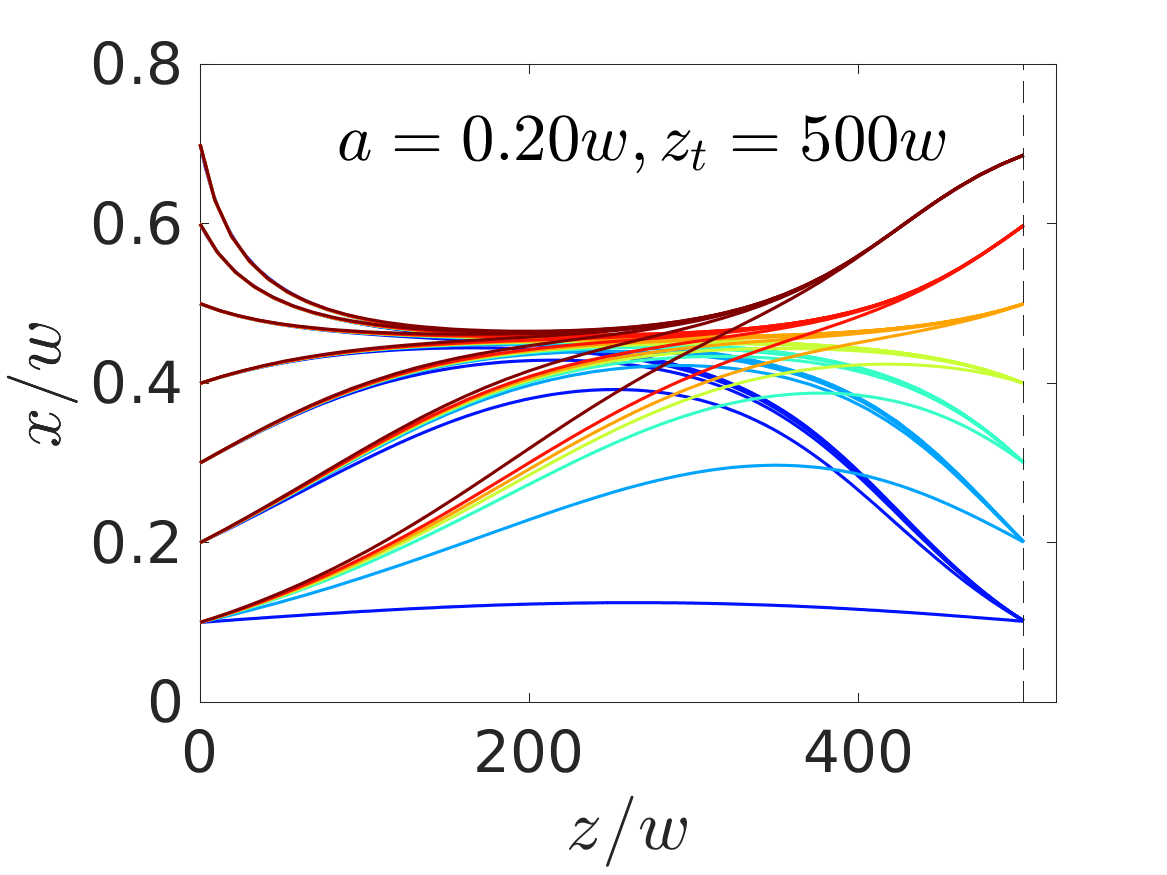}
\caption{Optimal control-force protocols (left) and 
particle trajectories in the $x$-$z$ plane (right) found for steering colloids with radius $a=0.2w$ from a set of initial positions to a set of targets,
which both assume the same values $\{0.1,0.2,0.3,0.4,0.5,0.6,0.7\}$. 
The same color refers to trajectories ending at the same target position $x_t$. The vertical dashed  line
indicates $z_t$.}
\label{fig:solutions_a20_500}
\end{figure}  

In Fig.~\ref{fig:solutions_a20_500} we plot the optimal force protocol (left) and the trajectories (right) resulting from different start and target positions using the particle radius $a=0.2w$ and channel length $z_t-z_i = 500w$. Interestingly, the regularization term 
generates solutions, where the force is zero at first, meaning that all colloids travel towards the equilibrium position at zero control force except when both $x_i$ and $x_t$ are close to the channel center. 
In particular, for $x_i=x_t=0.1w$, the algorithm chooses a nearly constant force protocol,
clearly recognizable in Fig.\ \ref{fig:solutions_a20_500}, left. In all other cases, the control force increases or decreases monotonously
starting around $t=300 w^2\nu^{-1}$, 
which corresponds to a traveled distance
between $z=250w$ and $300w$.
At the end of the trajectory the target is reached with high precision at $z=z_t$.

Note, the control-force protocols are similar for the same target position (same color).
However, the increase/decrease from zero 
starts earlier if the initial lateral position is closer to the channel center.
This is because flow velocity is larger and thus particles travel faster downstream.

\begin{figure}
	\centering
	\includegraphics[width=0.40 \textwidth]{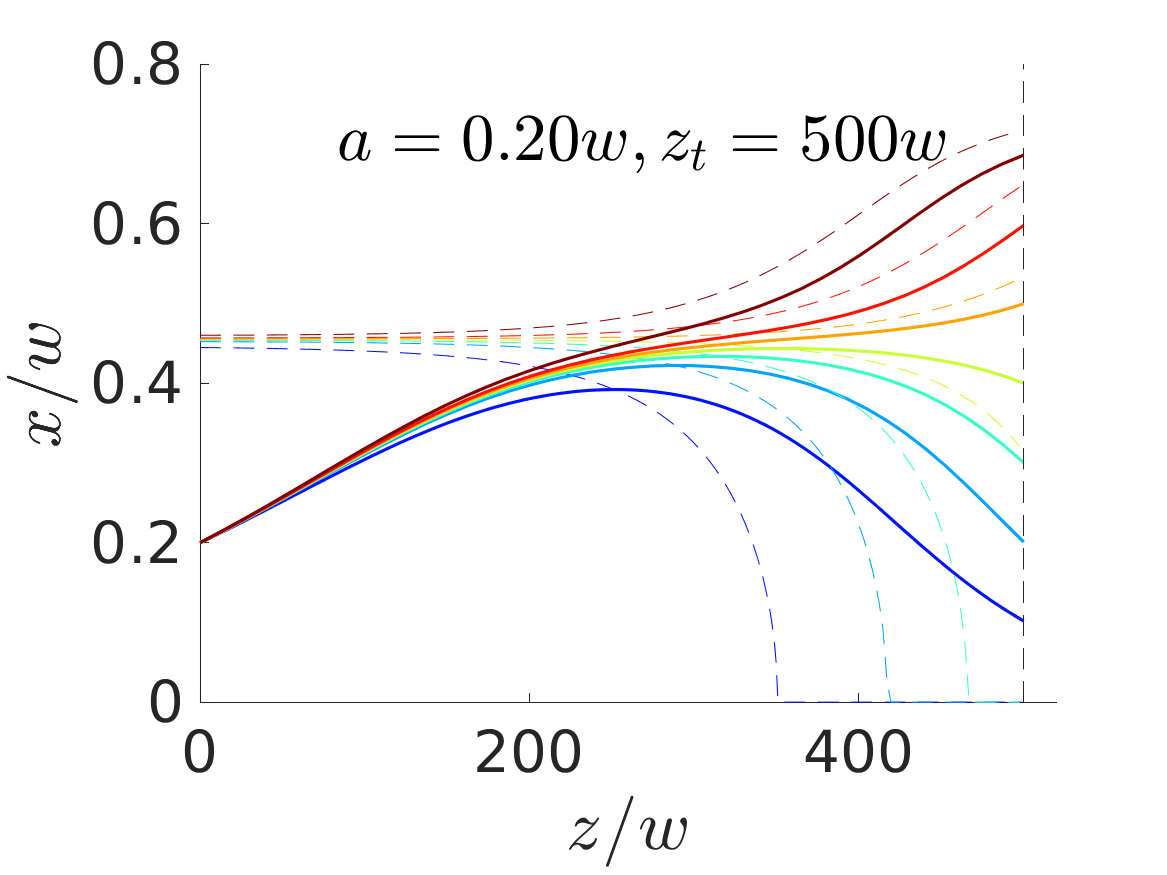}
	\caption{Solid lines: 
	Optimal particle trajectories in the $x$-$z$ plane found for steering colloids with radius $a=0.2w$
	from the initial position $x_i=0.2w$ to a set of targets $\{0.1,0.2,0.3,0.4,0.5,0.6,0.7\}$. 
	Dashed lines: Sequence of fixed-point positions of the lift-force profiles resulting from the optimal control-force protocols $f_{\mathrm{ctl}}(t)$ (same colors represent the same force protocol).}
	\label{fig:solutions_a20_500_attractors}
\end{figure}  

From the explanation so far, one could assume that the particle instantaneously follows the stable fixed points of the
lift-force profiles associated with the optimal control-force protocol $f_{\mathrm{ctl}}(t)$.
We plot the sequence
of fixed points as dashed lines in Fig.\ \ref{fig:solutions_a20_500_attractors} in the $x,z$ plane together with the realized particle trajectories starting at $x_i=0.2w$ and ending at different target positions. Clearly, the particle does not follow the sequence of stable fixed points, since migrating there is hindered by viscous friction.
For target positions close to the channel center, the control force induces a fixed point at the center ($x=0$) for the last part of the trajectory to accomplish the 
trajectories bend downward in Fig.~\ref{fig:solutions_a20_500_attractors}.

\subsubsection{Comparison with constant force strategy}

\begin{figure}
	\centering
	\includegraphics[width=0.49\textwidth]{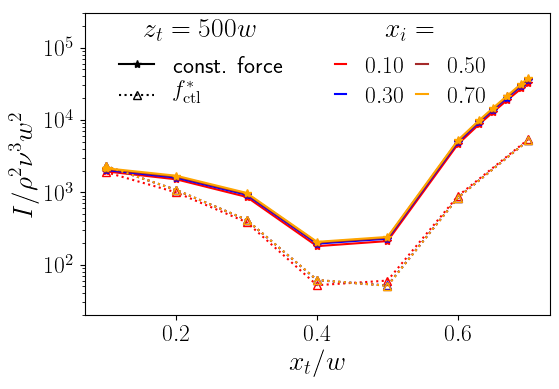}
	\caption{Comparison between the constant-force strategy (solid lines) 
	and the optimal control-force strategy (dashed line), for different 
	inital lateral 	positions and a channel length $z_t=500 w$. We show a
	semi-logarithmic plot of the cost functional $I$ of the axial control force versus  lateral target position $x_t$.	
	}
	\label{fig:integral_compare}
\end{figure}

We compare the costs of the constant-force strategy 
and the optimal control-force scheme using the cost functional $I:=\int_{0}^{T^*} \vert f^*_\mathrm{ctl}\vert^2(t)dt$. It integrates the 
square of the control force along the particle trajectory,
where lateral initial and target positions are equal for both strategies.
Note that $I$ is a measure for the energy costs needed to realize the control schemes. To compare both strategies, we decided to work with a constant channel length $z_t$ as a typical situation in experiments.
While the optimal-control scheme can be adjusted to such a specific axial target $z_t$, for the constant-force strategy the necessary
channel length varies depending on the lateral target position, as we discussed in Sect.\ \ref{sec:constant_force}.
In Fig.\ \ref{fig:integral_compare}, left we plot the cost functional $I$ for both strategies versus target position $x_t$
for different initial lateral positions $x_i$. The channel length is always $z_t=500w$.
Interestingly, the curves for each strategy are all very similar and therefore independent of $x_i$. 
The reason is the control-force profiles for different $x_i$ but same $x_t$ in Fig.\ \ref{fig:solutions_a20_500} have all very 
similar shape and are mostly shifted relative to each other. Clearly, the optimal-control scheme is less costly than the constant-force strategy up to an order of magnitude, besides for the smallest target position
$x_t/w = 0.1$. Here, we note that for the constant-force strategy the channel length $z_t=500w$ is not sufficient for reaching 
targets with $x_t/w \le 0.4$. Taking into account the need for longer channels, the costs of the constant-force strategy goes up.
In contrast for large $x_t \ge 0.5$ a channel length smaller than $500w$ is sufficient, which reduces the costs. But those do not 
fall below the costs of the optimal-control scheme.

\subsubsection{Dependence on particle size}
\begin{figure}
\centering
\includegraphics[width=0.40 \textwidth]{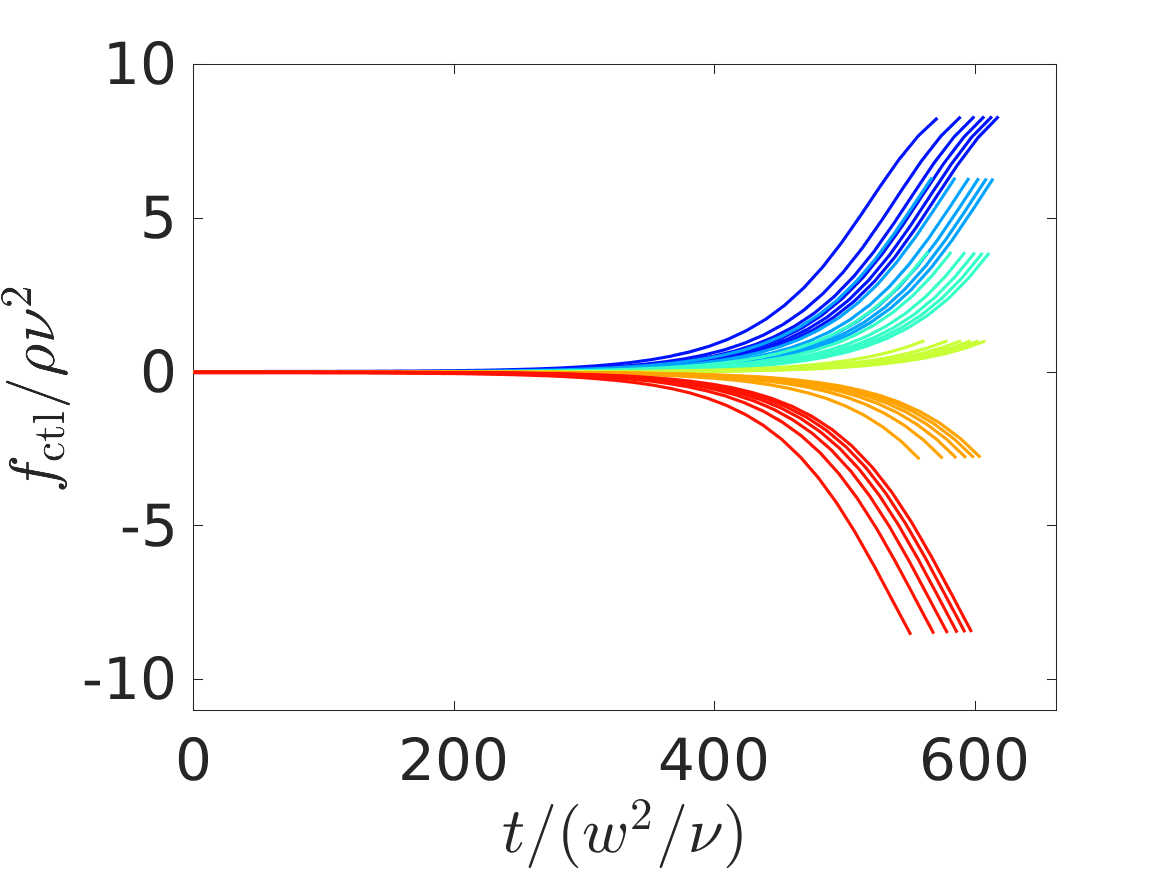}
\includegraphics[width=0.40 \textwidth]{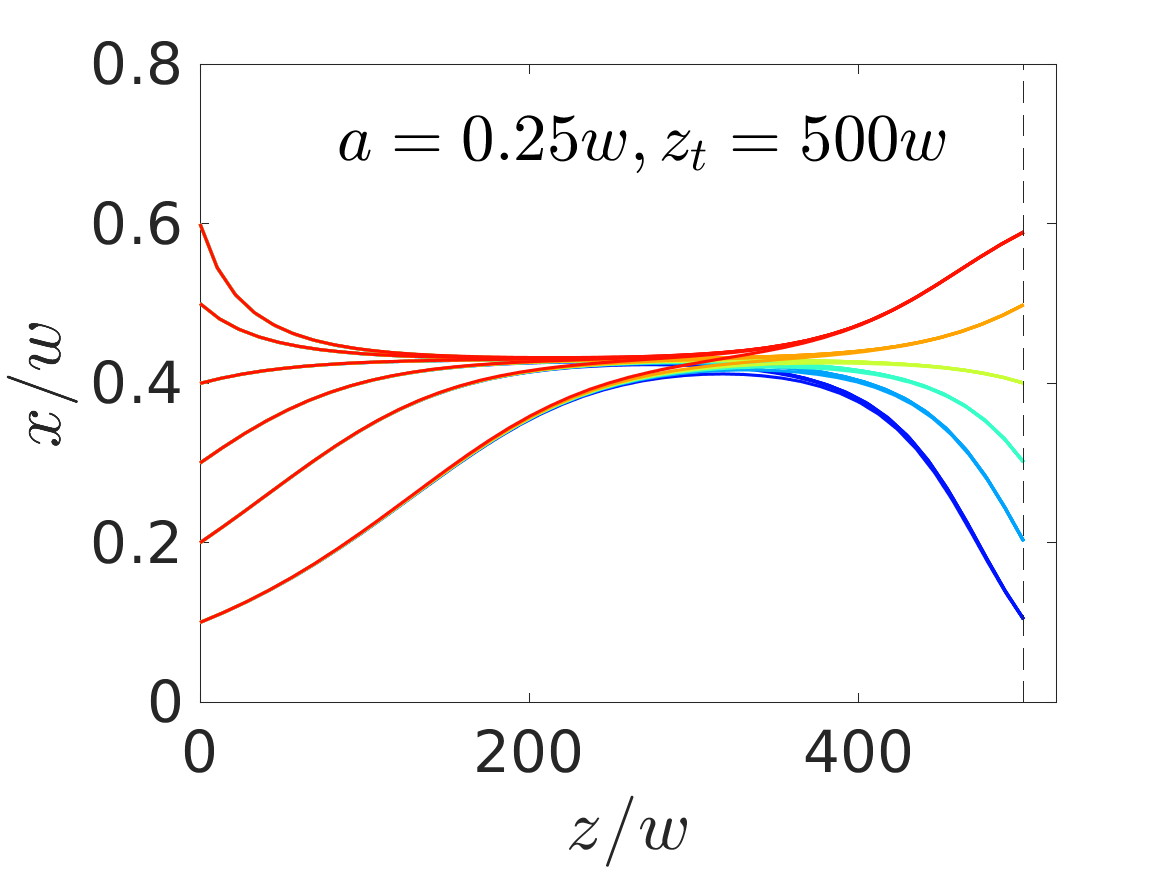}

\includegraphics[width=0.40 \textwidth]{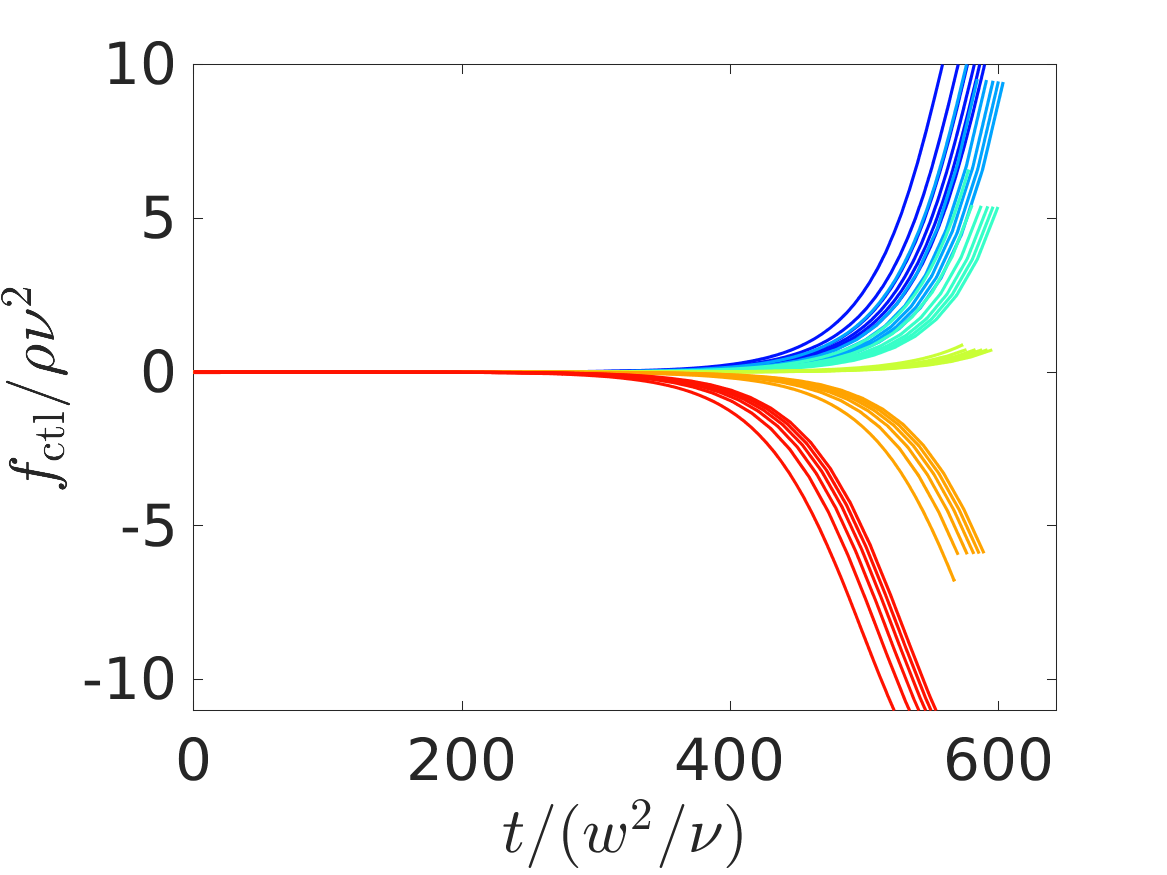}
\includegraphics[width=0.40 \textwidth]{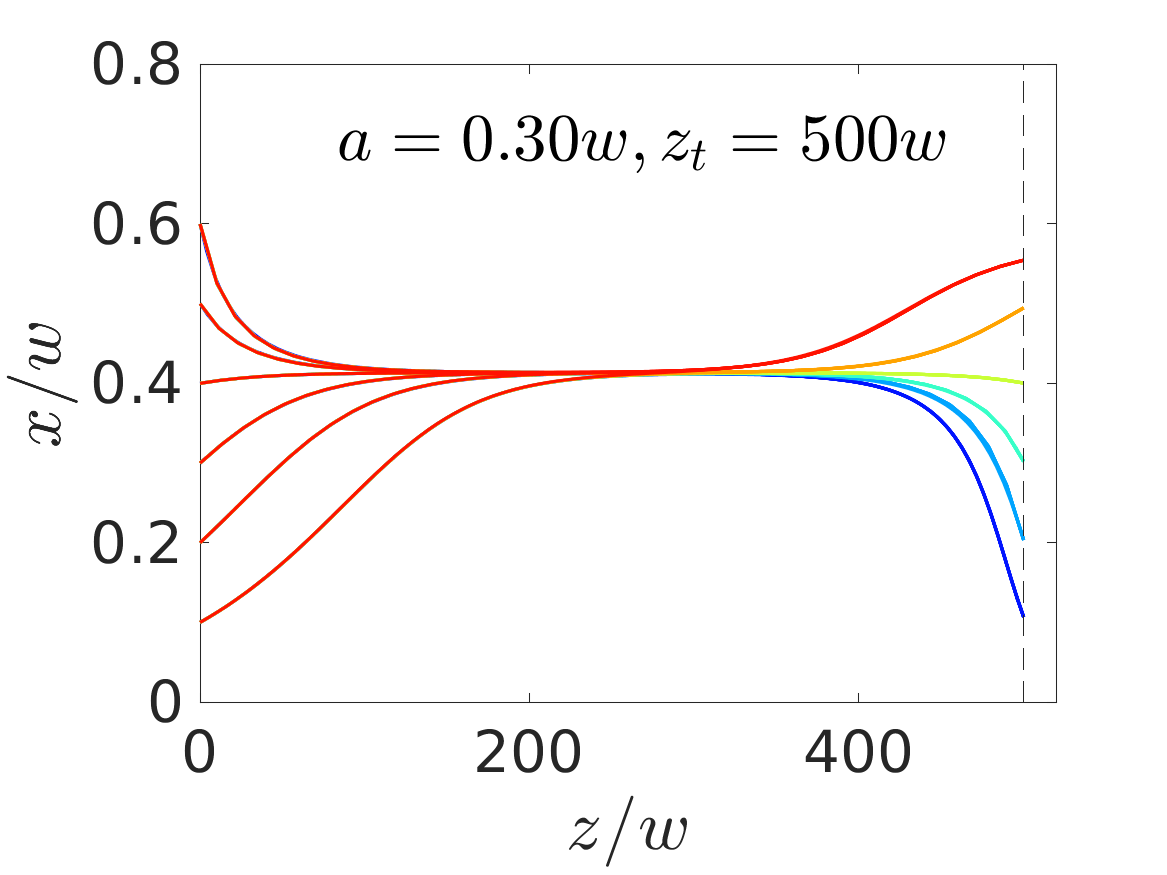}
\caption{Optimal control-force protocols (left column) and particle trajectories in the $x$-$z$ plane (right column) found for steering colloids with radius $a=0.25w$ (top row) and $a=0.3w$ (bottom row) from a set of initial positions to a set of targets, which both assume the same values $\{0.1,0.2,0.3,0.4,0.5,0.6\}$. 
The same color refers to trajectories ending at the same target position $x_t$. The vertical dashed  line
indicates $z_t$.
}
\label{fig:solutions_a25_a30}
\end{figure}  

Increasing the particle size, strongly increases the strength of the inertial lift force ($f_\mathrm{lift} \propto (a/w)^4$ for very small particles). Thus, the fixed-point position at the initially zero 
control force is reached faster, as a comparison of the force protocols and the partcle trajectories in Figs.\ \ref{fig:solutions_a20_500}
and \ref{fig:solutions_a25_a30} for different radii shows.
The control force remains longer at a zero value before it steers the particle to its target position. However, for the larger particles 
this then also requires larger control forces to steer them laterally because they experience a higher drag force and it is therefore 
harder to move them relative to the external flow. Furthermore, 
as before for target positions further away from the zero-force fixed-point position, larger control forces are necessary for steering, which makes sense. Finally, as we already noted in Sect.~\ref{sec:saffman}, since the Saffman force ($f_\mathrm{S} \propto a^2$) grows less strongly with the radius than the focussing inertial lift force, one again needs larger control forces for particle steering, which also drives up the whole costs.
Thus, in all trajectories the particles utilize inertial focussing at first
to relax towards the zero-force equilibrium positions and then the control force is switched on.

\begin{figure}[t!!]
\centering
\includegraphics[width=0.40 \textwidth]{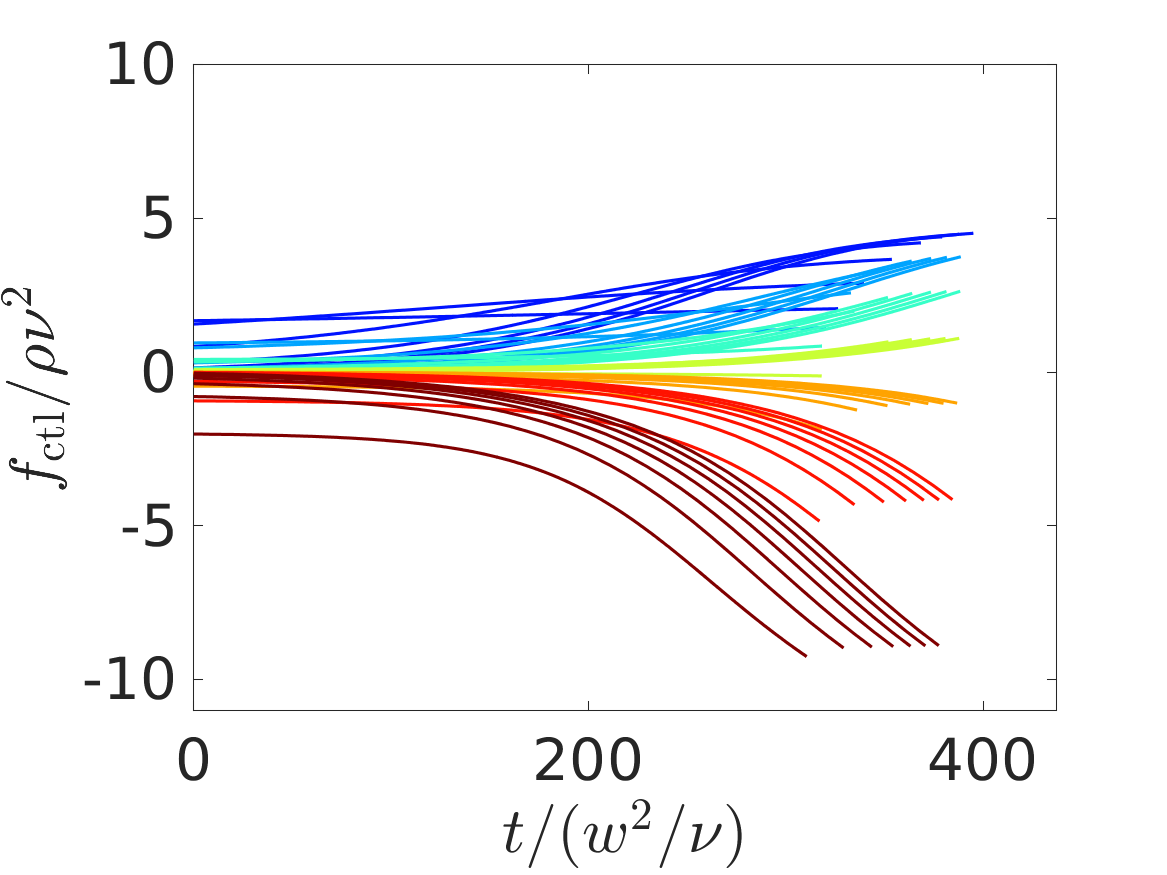}
\includegraphics[width=0.40 \textwidth]{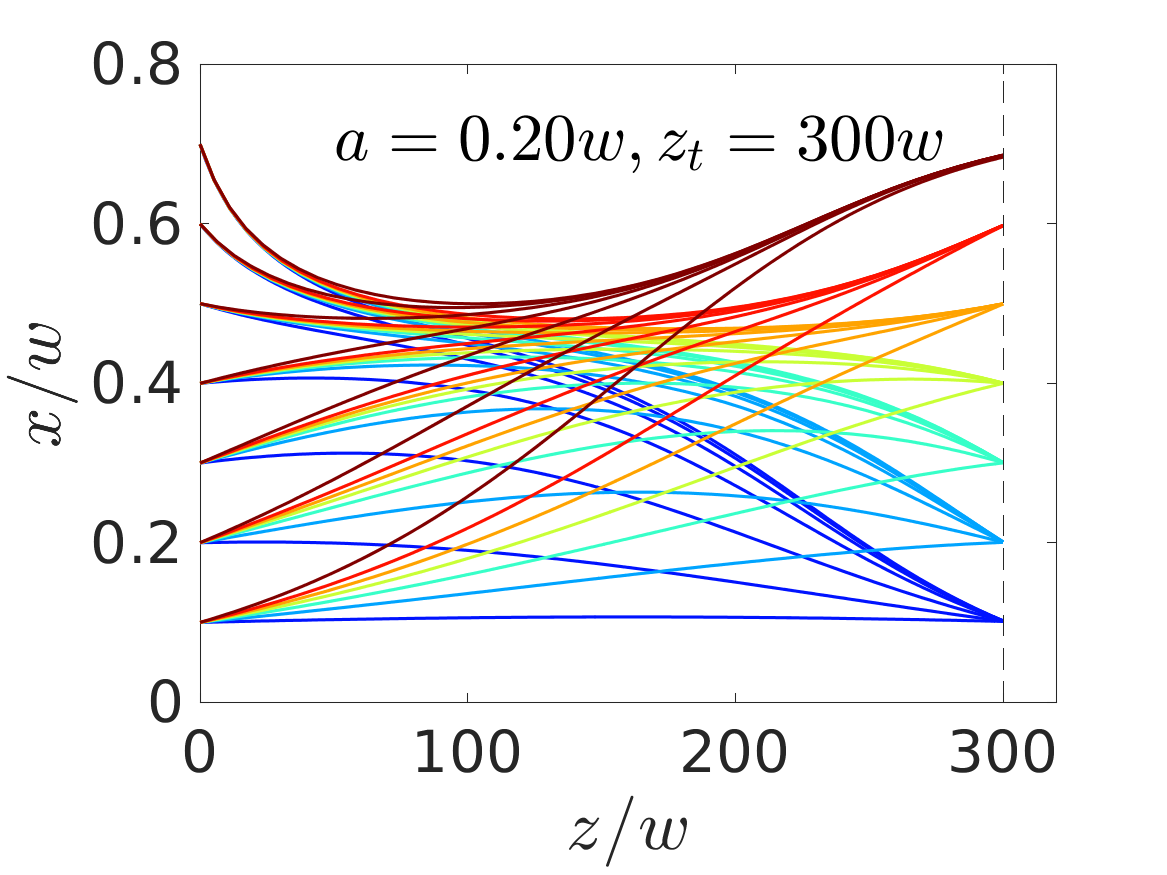}

\includegraphics[width=0.40 \textwidth]{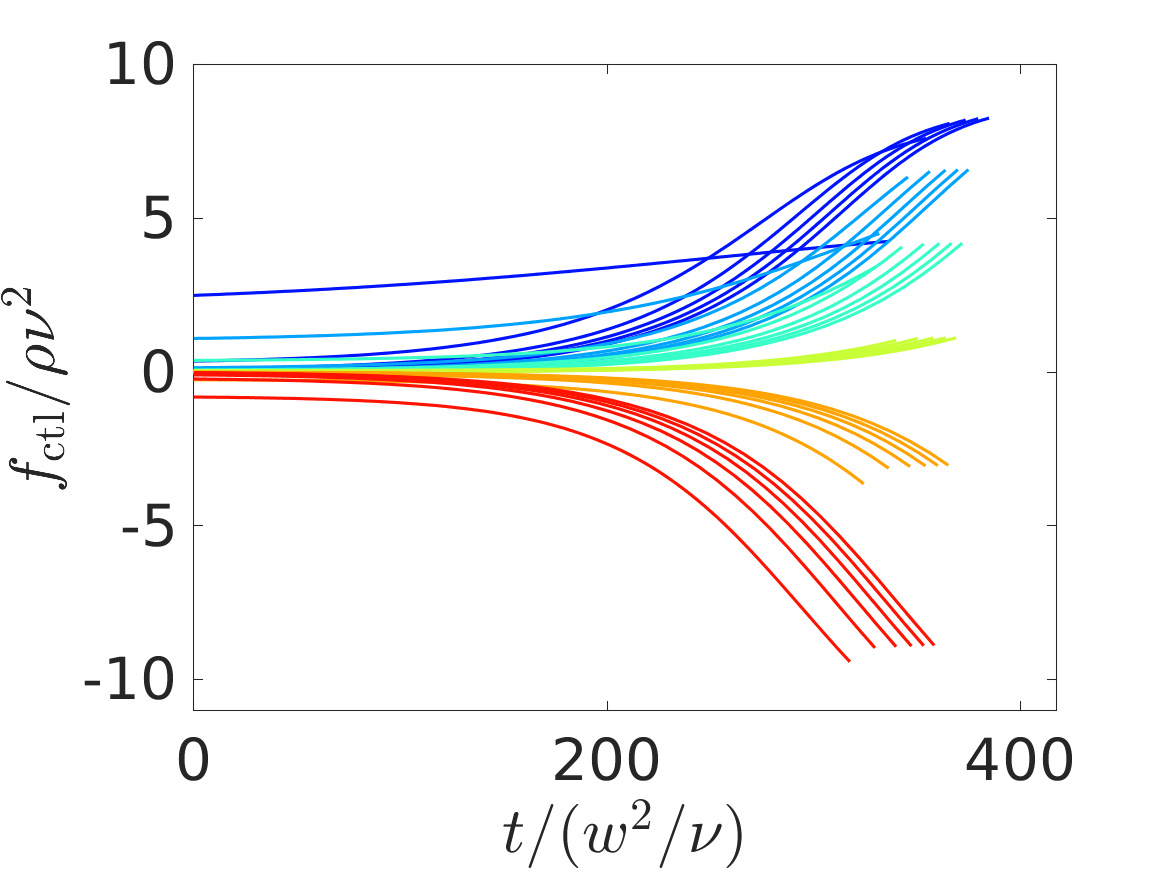}
\includegraphics[width=0.40 \textwidth]{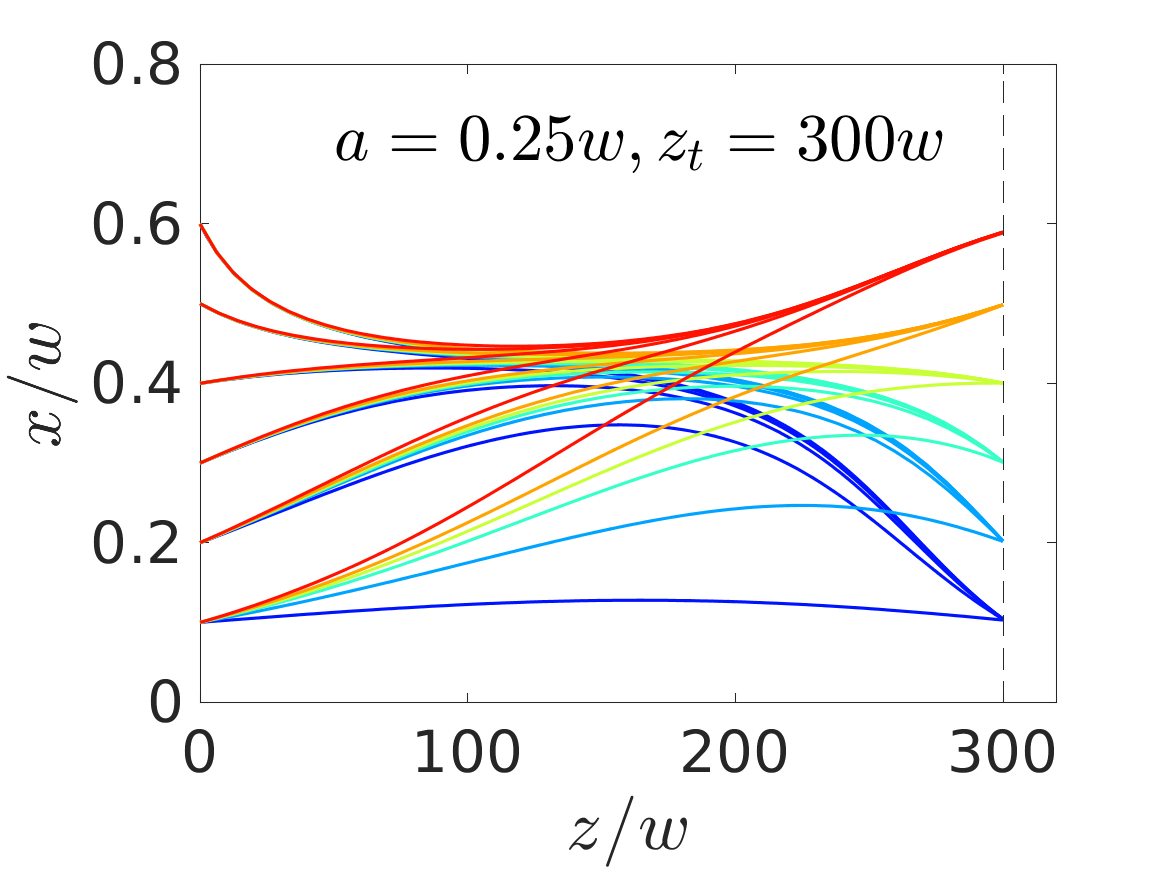}

\includegraphics[width=0.40 \textwidth]{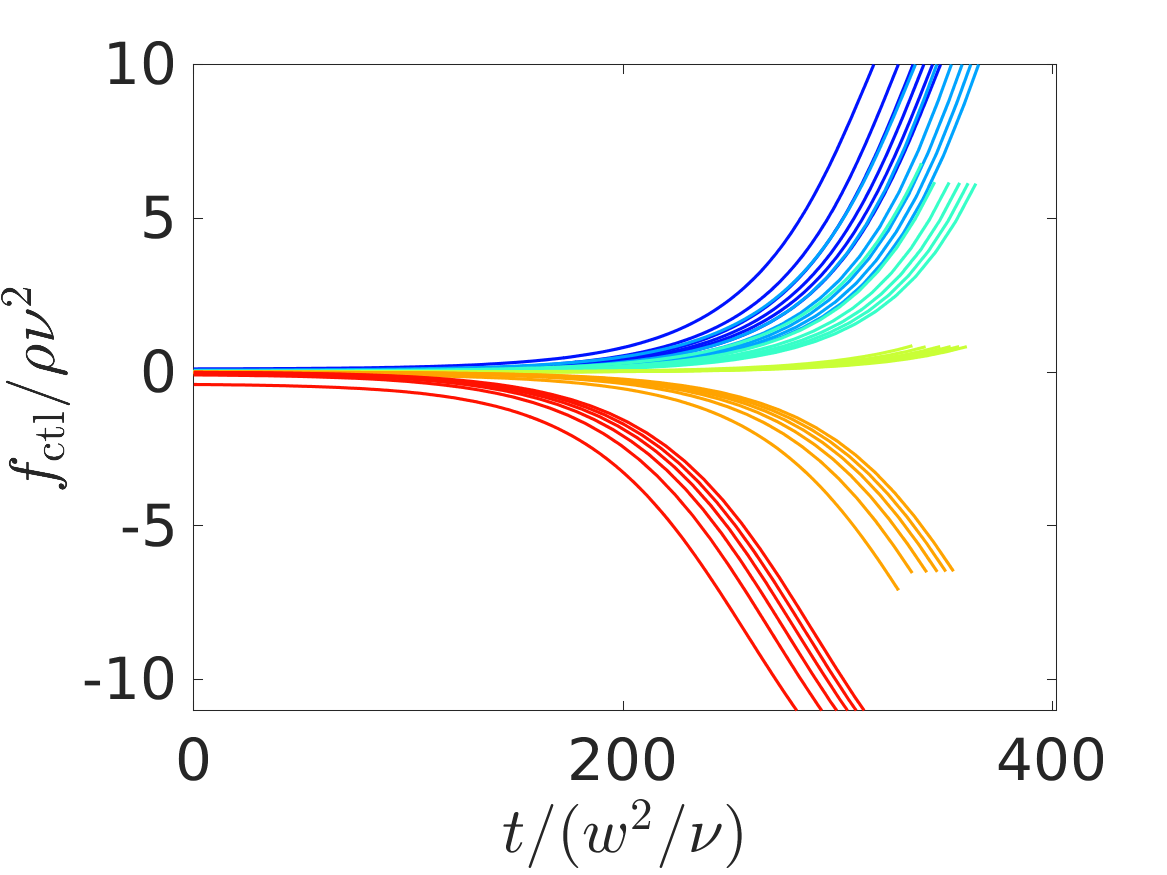}
\includegraphics[width=0.40 \textwidth]{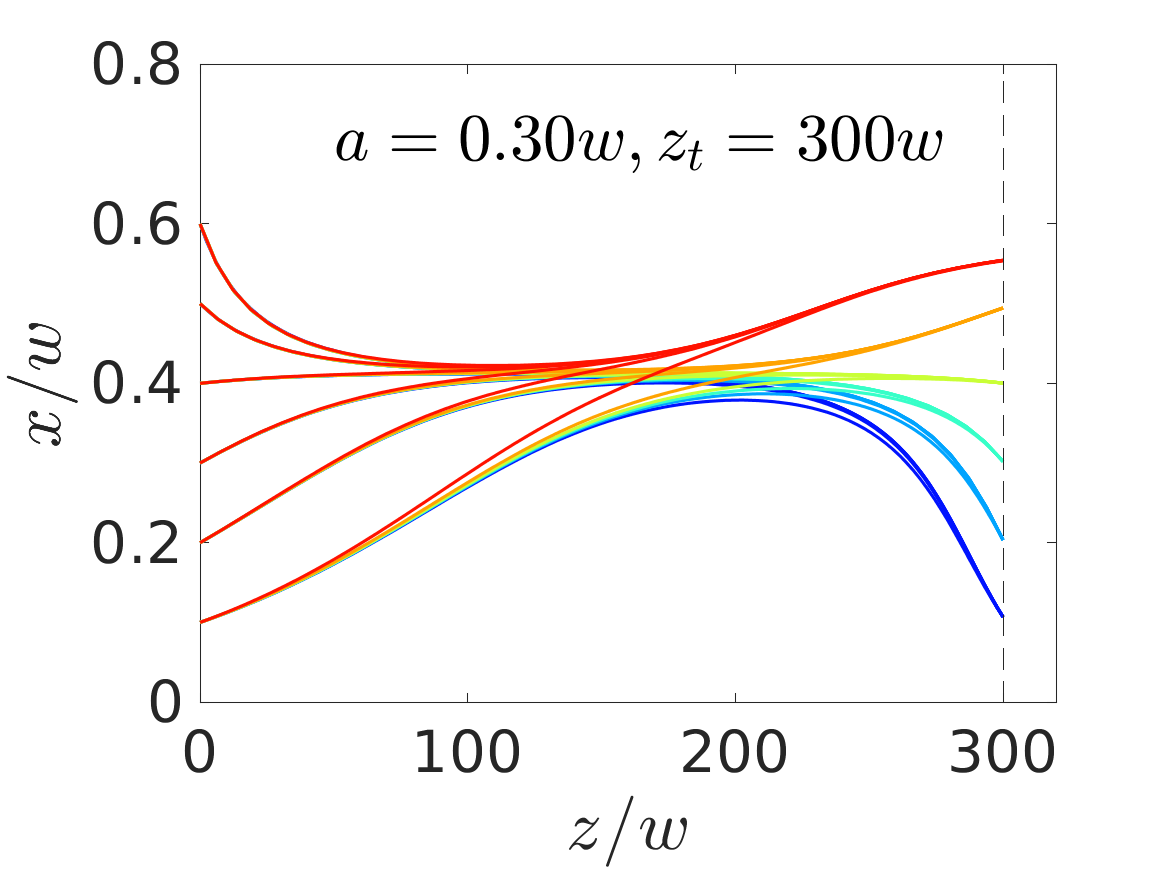}
\caption{Optimal control-force protocols (left column) and particle trajectories in the $x$-$z$ plane (right column) found for steering colloids with radii $a/w=0.2$, $0.25$, and 0.3 (top, middle, and bottom row) to an axial target position $z_t = 300 w$. The vertical dashed  line indicates $z_t$.}
\label{fig:solutions_z300}
\end{figure}  

\subsubsection{Dependence on channel length}
Since the control forces obtained in Fig.~\ref{fig:solutions_a20_500} and \ref{fig:solutions_a25_a30} all remain at zero at the 
beginning of the particle trajectories, it should be possible to further decrease the channel length. Indeed, we managed to obtain numerically stable solutions for a channel length of $z_t = 300w$, which we show in Fig.~\ref{fig:solutions_z300}. Here, the trajectories do not ($a/w = 0.2$ and $0.25$) or only shortly ($a/w = 0.3$) stay on the lateral focus position, and thus the control force is always non-zero or zero for a short time. Interestingly, the algorithm chooses relaxation towards the equilibrium position for the largest radius $a=0.3w$. Again, the inertial lift force 
increases strongly with the particle radius and,
therefore, it is too costly to compete against it 
with a non-zero control force over the whole simulation time. Instead, the algorithm chooses to drive up the control force to high absolute values but for a shorter time period at the end of the trajectory.

\subsection{Controlled steering of multiple colloids}
We use our model to steer two or more particles to their respective targets. As a first approximation we neglect here two-particle interactions. It is known that the lift-force profiles of two particles are influenced due to secondary flows, when their axial distance is smaller or of the order of the channel width~\cite{AminiDiCarlo2014,SchaafStark2019}.
In the following, we first consider the steering of a pulse of equal-sized particles, and then investigate
the lateral separation of two particles with different radii under the same control-force protocol.

\subsubsection{Steering a pulse of colloids}
\label{sec.pulse}

\begin{figure}
	\centering
	\includegraphics[width=0.49\textwidth]{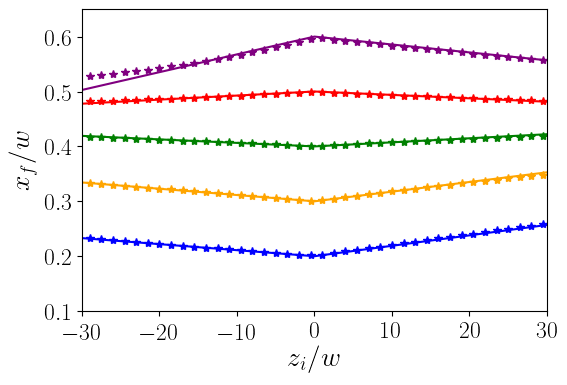}
	\includegraphics[width=0.49\textwidth]{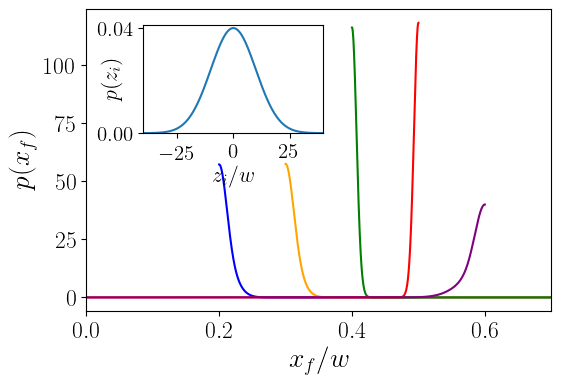}
	\caption{Left:
	Final lateral position $x_f$ for particles with initial axial position $z_i$ when steered by the axial control force
	$f_\mathrm{ctl}^*(t)$, which is opimized for steering the central particle with $z_i = 0$ and $x_i = 0.2$ to different
	target positions $x_t/w = 0.2$, $0.3$, $0.4$, $0.5$, and $0.6$ at the axial target position $z_t=500w$.
	The linear fits by solid lines are hardly visible. Right: Distributions of final lateral positions $x_f$ for the Gaussian distribution of initial axial positions $z_i$ shown in the inset.
The colors indicate different target positions $x_t/w$ as in the plot on the left.
}	
	\label{fig:pulse_endpositions}
\end{figure}

In the following we consider the situation where a pulse of colloids is injected at the inlet of a microchannel. We assume they all have the same initial lateral position $x_i$ but are spread along the axial direction according to a Gaussian distribution as shown in the inset of Fig.\ \ref{fig:pulse_endpositions}, right. At time $t=0$ the center of the Gaussian is at $z_i =0$.
Now, we ask which final lateral position $x_f$ the colloids attain when reaching the axial target position $z_t$ under the action 
of the optimal control force $f_\mathrm{ctl}^*(t)$.
For the latter we use the optimal force protocol calculated for the central initial position at $z_i=0$. It is switched on at $t=0$ and switched off at the time  $T^*$ when the particle 
moving on the original optimized trajectory with $z_i=0$ has reached the axial target position $z_t$.
Since all particles start on the same initial lateral position $x_i$ at $t=0$, they move on replicas of the opimized trajectory but shifted along the $z$ direction by $z_i$. Colloids with $z_i >0$ precede the original optimized trajectory and therefore experience the control force until they have reached 
$z_t$. However, for colloids lagging behind the 
optimized trajectory ($z_i <0$), we simply turn off the control force once the optimized time period $T^*$ has passed and wait until they have reached 
$z=z_t$. During this time, the lateral motion is completely determined by inertial focusing 
without any Saffman force, where the focusing position is $x_\mathrm{eq}^0$. This allows for two scenarios: If the lateral target position $x_t$ is closer to the channel center than the focusing position ($x_t < x_\mathrm{eq}^0$), we know from Sec.~\ref{sec:control_single} that the optimal trajectory approaches the target from above. Therefore, both the preceding and lagging colloids 
will reach a final lateral position $x_f > x_t$.
In contrast, for targets closer to the channel wall than the focusing position ($x_t  > x_\mathrm{eq}^0$), both leading and lagging colloids end up closer to the center than the target ($x_f < x_t$). 

The qualitative description is confirmed by Fig.\ \ref{fig:pulse_endpositions}, left , where we plot the final lateral positions $x_f$ versus $z_i$ for different target positions $x_t$. 
They follow piecewise linear functions, where the two arms have different slopes since different mechanisms determine the final positions of preceding and lagging colloids.
Only for the target position $x_t = 0.6$ do the final positions $x_f$ deviate from the linear course for very negative $z_i$.
We note that the deviations of $x_f$ from the target remain small even when the axial particle positions are spread over $30w$ 
to both sides of $z_i = 0$.

Taking the Gaussian distribution of initial axial positions in the inset of Fig.\ \ref{fig:pulse_endpositions}, right with standard deviation
$\sigma_z=10w$ and assuming the linear dependence $x_f(z_i)$ for the final lateral positions, one can readily write the
distribution $p(x_f)$ of final lateral positions. It is a superposition of two Gaussian functions, where only one half is used from 
each Gaussian (see below). The resulting distributions for the different target positions are presented in 
Fig.\ \ref{fig:pulse_endpositions}, right. Although the axial width of the initial distribution is ca. $50w$, the final positions only 
deviate a little from the target position. The distribution $p(x_f)$ for $x_t = 0.4 w$ (green curve) is sharpest since $x_t = 0.4 w$ 
is closest to the zero-force focusing position $x_\mathrm{eq}^0$. The distributions become broader when $x_t$ is moved
towards the wall or the channel center, respectively. Thus we demonstrate here, that a pulse of colloidal particles fairly
spread in the axial direction can be focussed into one target position at the channel outlet using one control-force protocol 
for all the particles.

At the end we shortly present the derivation of the distribution $p(x_f)$ of final positions at the channel outlet.
Since particles with a specific initial axial position $z_i$ move to a specific $x_f$, one can derive $p(x_f)$ directly from the distribution $p_z(z_i)$ and obtain:
\begin{equation}
p(x_f) = p_z(f^{-1}_{+}(x_f))\vert (f^{-1}_{+})'(x_f)\vert +  p_z(f^{-1}_{-}(x_f))\vert (f^{-1}_{-})'(x_f) \vert.
\end{equation}
Here, $x_f=f_\pm(z_i) = a_\pm z_i + x_t$ is the piecewise linear function from fitting the curves in Fig.\ \ref{fig:pulse_endpositions}, 
left, $z_i=f^{-1}_{\pm}(x_f)$ is its inverse function, and $(f^{-1}_{\pm})'(x_f) = 1/a_\pm$. Taking a Gaussian distribution for $p_z(z_i)$, the final distribution $p(x_f)$ is a sum of two shifted and rescaled Gaussians with means at $x_t$. However, since the value range of $f_\pm$ is either $(-\infty,x_t]$ or $[x_t,\infty)$, the end result is a sum of two half-normal distributions, either to the left ($x_t > x_\mathrm{eq}$) or to the right ($x_t < x_\mathrm{eq}$) of the 
mean $x_t$ of the full Gaussian.
This is readily seen in Fig.~\ref{fig:pulse_endpositions}, right. 
{To have a quantitative measure for the width of the distribution $p(x_f)$, we calculate its mean value:
\begin{equation}
\mu = x_t
\pm
\frac{\sigma_z}{\sqrt{2\pi}}(\vert a_+\vert +\vert a_- \vert),
\end{equation}
where the plus sign applies to $x_t < x_\mathrm{eq}^0$ and vice versa.
The deviation from $x_t$ provides a measure for the width of $p(x_f)$. It is determined by the slopes $a_{\pm}$
of the linear fits to $x_f=x_f(z_i)$. Since they are small also the width is small and it decreases when 
$x_t$ approaches $x_\mathrm{eq}^0$.
Ultimately this small width comes from the fact that drift velocities in lateral channel direction are much smaller than
the axial flow velocity. This means, inertial transport is much weaker than axial transport due to Poiseuille flow.

\subsubsection{Separation of Particles}
\label{sec.separation}

\begin{figure}
	\includegraphics[width=0.49\textwidth]{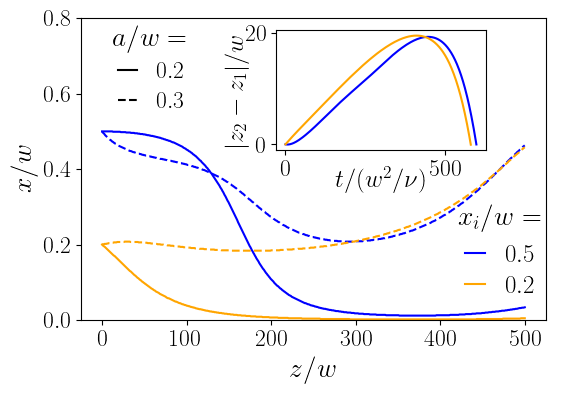}
	\includegraphics[width=0.49\textwidth]{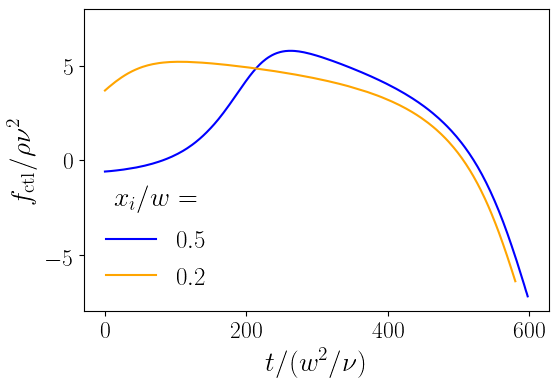}
	\caption{Maximizing the lateral 
	distance of two particles with different radii $a/w= 0.2$ and $0.3$
      using the same control-force protocol $f_\mathrm{ctl}^*(t)$.
      The particles both enter at the same inlet at $x_i$ and travel an axial distance with optimal value $z_t=500w$ during time $T^*$.
      Left: Trajectories of the two particles for both initial conditions. Inset: Axial separation over time of the two particles.
      Right: Optimal control-force protocols for the two initial lateral positions $x_i=0.2w$ and $0.5w$.}
	\label{fig:separation}
\end{figure}

In the end we examine the case where two particles of different size are steered using the same control-force protocol. 
\fr{Thus, we assume here that both particle types experience the same external force independent of their sizes.}
From Fig.~\ref{fig:stable_fixed_points} we already know that this is possible: A properly chosen constant axial force can drive the smaller particle to the center while the larger particle stays at a finite distance from the center. Here we aim to maximize the lateral distance
after both particles have traveled the distance $z_t$ in axial direction. At the end of Sect.\ \ref{sec:system_control} we already formulated the appropriate cost functional for maximimizing the lateral distance between both particles. 
In Fig.~\ref{fig:separation} we show the resulting trajectories (left) and force protocols (right) for two particles with
radii $a_1=0.2w$ and $a_2=0.3w$ and the axial target $z_t=500w$. Without control force, these particles would arrive at very similar positions, because their zero-force equilibrium positions are very close to each other. We 
present results for two cases where both particles start at the same initial position either at $x_i=0.2w$ or $x_i=0.5w$. Again, we assume they do not interact. Interestingly, the 
control-force protocols for both cases look rather different in the beginning. However, in both cases the smaller particle (solid lines in Fig.~\ref{fig:separation}, left) is pushed towards the centerline, 
while the larger particle (dashed lines) moves towards the channel wall during the second halves of the trajectories. The separation reached at the end is $\Delta x = 0.45w$ for $x_i=0.2w$ and $\Delta x= 0.43w$ for $x_i=0.5w$, which is not attainable with any passive method.
 
\begin{figure}
	\centering
	\includegraphics[width=0.49\textwidth]{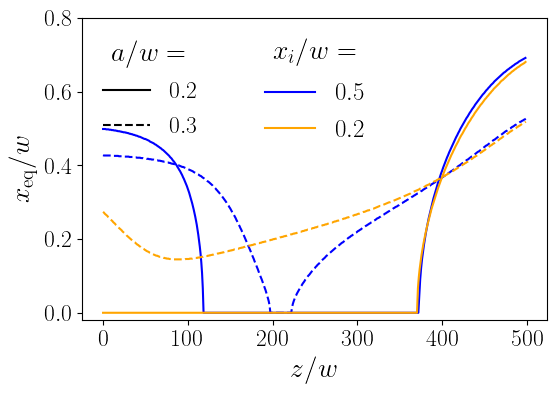}
	\caption{Instantaneous stable fixed points resulting from the applied forces in Fig.~\ref{fig:separation}(left).
	}
	\label{fig:separation_axial_fp}
\end{figure}

To develop a better understanding for the optimal control-force protocols of Fig.\ \ref{fig:separation}, right, we show in Fig.~\ref{fig:separation_axial_fp} the momentary stable fixed points for the smaller and larger particles corresponding to the momentary 
axial control force, when the particles are at position $z$. The path of the momentary fixed points reflects the particle trajectories of Fig.\ \ref{fig:separation}, left, where the algorithm attempts to steer the smaller particle (solid lines) to the channel center and 
the larger particle (dashed lines) towards the wall. For 
initial position $x_i=0.5w$ (blue lines) the fixed point at zero control force is 
closer to the channel center (around $0.4w$),
therefore the control force close to zero is sufficient to move, in particular, the smaller particle towards the center. Then it rises noticeably, bringing the smaller particles
to the center as documented by the course of the momentary fixed point.
In contrast, the initial position $x_i=0.2w$ (orange color)
is already closer to the channel center. Thus the control force is switched on immediately to push the smaller particle (orange solid line)
to the channel center, where the momentary fixed point is located. 
As we know from Fig.\ \ref{fig:stable_fixed_points}, left, the fixed point of the large particle does not react so strongly to the control force. It is only shifted towards the center but hardly reach it. Nevertheless, this initial behavior causes the minima in the dashed trajectories of Fig.~\ref{fig:separation}, left.
Then in both cases the axial control force identified by our algorithm becomes strongly negative and the momentary fixed 
points are pushed towards the wall, even stronger for the smaller particles. Nevertheless, since the lift force scales with the
particle radius $a^2$, the larger particles are pushed towards the wall, while the smaller particles stay close to the channel 
center and hardly move away from it (see Fig.~\ref{fig:separation}, left).

\section{Conclusions}
\label{sect.concl}

We applied concepts from optimal control theory to a setup from inertial microfluidics and managed to precisely steer single particles from 
a microchannel inlet to an outlet using a time-dependent axial force,
which controls the lateral inertial lift force via the Saffman effect. Our results show that the optimal control force exploits conventional inertial migration since in the beginning it is zero so that the particle drifts towards its lateral equilibrium position. Only then
the control force is switched on so that the particle is pushed towards a target position. Due to this property steering with an optimized control force is cheaper than a strategy where a constant axial force is used for steering. 
Additionally, the optimal-control strategy can be implemented for different channel lengths, which makes this approach versatile. 
We also used the optimal control-force protocol for a single particle to demonstrate that even a pulse of particles spread 
along the channel axis can be steered to a target with only a small spread around the exact target position. Finally, we showed 
how a single optimized control-force protocol can separate particles of similar radii $a_1=0.2w$ and $a_2=0.3w$.
The lateral distances reached for a channel length of $500w$ 
and different initial positions are considerably larger than a passive strategy could achieve.

\fr{It would be interesting to explore different channel geometries in the future, such as rectangular cross sections with different aspect ratio or triangular cross sections \cite{KimLee2016}, because they strongly influence the locations of the fixed points of the inertial lift force. 
For triangular microchannels this would require to determine the lift-force profile in the cross-sectional plane and
not just on one axis.
}

We consider particle steering by an optimal axial control force as an innovative method for \fr{targeting precise 
positions at the channel outlets, which will then have implications for particle separation and filtration. We} hope that our work stimulates \fr{future} efforts towards an experimental realization. \fr{As we outlined above, to realize the axial control force, we suggest the use of electric fields in combination with electrophoresis, which has already been applied in experiments on the Saffman effect and
inertial migration \cite{KimYoo2009,YuanLi2016} and also studied in theory\ \cite{ChoudharyPushpavanam2019}.
Typically, micron-sized particles can exhibit electrophoresis and it is also realized for biological cells \cite{KorohodaWilk2008}.
}

\vspace{6pt} 



\authorcontributions{
Conceptualization, F.R., C.S., and H.S.; methodology, F.R., C.S., and H.S.; software, F.R, and C.S..; validation, F.R., C.S., and H.S.; formal analysis, F.R., C.S., and H.S.; investigation, F.R.; resources, F.R., and C.S.; data curation, F.R., and C.S.; writing--original draft preparation, F.R, and H.S.; writing--review and editing, F.R.,C.S., and H.S.; visualization, F.R., and H.S.; supervision, H.S.; project administration, H.S.; funding acquisition, H.S. 
All authors have read and agreed to the published version of the manuscript.
}

\funding{
This research was funded by Deutsche Forschungsgemeinschaft through the priority program SPP 1726 (grant number STA352/11) and through the collaborative research center SFB 910.
}

\acknowledgments{We thank F.~Tr{\"o}ltzsch \fr{and A.~Resing} for fruitful discussions.}

\conflictsofinterest{The authors declare no conflict of interest. The funders had no role in the design of the study; in the collection, analyses, or interpretation of data; in the writing of the manuscript, or in the decision to publish the results.} 

%

\appendixtitles{yes} 
\appendix

\reftitle{References}


\bibliography{lit}

\begin{thebibliography}{-------}
\providecommand{\natexlab}[1]{#1}

\bibitem[Squires and Quake(2005)]{SquiresQuake2005}
Squires, T.M.; Quake, S.R.
\newblock Microfluidics: Fluid physics at the nanoliter scale.
\newblock {\em Reviews of modern physics} {\bf 2005}, {\em 77},~977.

\bibitem[Ismagilov \em{et~al.}(2001)Ismagilov, Ng, Kenis, and
  Whitesides]{IsmagilovWhitesides2001}
Ismagilov, R.F.; Ng, J.M.; Kenis, P.J.; Whitesides, G.M.
\newblock Microfluidic arrays of fluid- fluid diffusional contacts as detection
  elements and combinatorial tools.
\newblock {\em Analytical chemistry} {\bf 2001}, {\em 73},~5207--5213.

\bibitem[Whitesides(2006)]{Whitesides2006}
Whitesides, G.M.
\newblock The origins and the future of microfluidics.
\newblock {\em Nature} {\bf 2006}, {\em 442},~368--373.

\bibitem[Weibel and Whitesides(2006)]{WeibelWhitesides2006}
Weibel, D.B.; Whitesides, G.M.
\newblock Applications of microfluidics in chemical biology.
\newblock {\em Current opinion in chemical biology} {\bf 2006}, {\em
  10},~584--591.

\bibitem[Di~Carlo(2009)]{DiCarlo2009}
Di~Carlo, D.
\newblock Inertial microfluidics.
\newblock {\em Lab on a Chip} {\bf 2009}, {\em 9},~3038--3046.

\bibitem[Amini \em{et~al.}(2014)Amini, Lee, and Di~Carlo]{AminiDiCarlo2014}
Amini, H.; Lee, W.; Di~Carlo, D.
\newblock Inertial microfluidic physics.
\newblock {\em Lab on a Chip} {\bf 2014}, {\em 14},~2739--2761.

\bibitem[Segr{\'e} and Silberberg(1961)]{SegreSilberberg1961}
Segr{\'e}, G.; Silberberg, A.
\newblock Radial particle displacements in Poiseuille flow of suspensions.
\newblock {\em Nature} {\bf 1961}, {\em 189},~209--210.

\bibitem[Sudarsan and Ugaz(2006)]{SudarsanUgaz2006}
Sudarsan, A.P.; Ugaz, V.M.
\newblock Multivortex micromixing.
\newblock {\em Proceedings of the National Academy of Sciences} {\bf 2006},
  {\em 103},~7228--7233.

\bibitem[Bhagat \em{et~al.}(2008)Bhagat, Kuntaegowdanahalli, and
  Papautsky]{BhagatPapautsky2008}
Bhagat, A.A.S.; Kuntaegowdanahalli, S.S.; Papautsky, I.
\newblock Enhanced particle filtration in straight microchannels using
  shear-modulated inertial migration.
\newblock {\em Physics of Fluids} {\bf 2008}, {\em 20},~101702.

\bibitem[Sajeesh and Sen(2014)]{SajeeshSen2014}
Sajeesh, P.; Sen, A.K.
\newblock Particle separation and sorting in microfluidic devices: a review.
\newblock {\em Microfluidics and nanofluidics} {\bf 2014}, {\em 17},~1--52.

\bibitem[Di~Carlo \em{et~al.}(2009)Di~Carlo, Edd, Humphry, Stone, and
  Toner]{DiCarloToner2009}
Di~Carlo, D.; Edd, J.F.; Humphry, K.J.; Stone, H.A.; Toner, M.
\newblock Particle segregation and dynamics in confined flows.
\newblock {\em Physical review letters} {\bf 2009}, {\em 102},~094503.

\bibitem[Lee \em{et~al.}(2010)Lee, Amini, Stone, and Di~Carlo]{LeeDiCarlo2010}
Lee, W.; Amini, H.; Stone, H.A.; Di~Carlo, D.
\newblock Dynamic self-assembly and control of microfluidic particle crystals.
\newblock {\em Proceedings of the National Academy of Sciences} {\bf 2010},
  {\em 107},~22413--22418.

\bibitem[Schaaf \em{et~al.}(2019)Schaaf, R{\"u}hle, and Stark]{SchaafStark2019}
Schaaf, C.; R{\"u}hle, F.; Stark, H.
\newblock A flowing pair of particles in inertial microfluidics.
\newblock {\em Soft matter} {\bf 2019}, {\em 15},~1988--1998.

\bibitem[Bretherton(1962)]{Bretherton1962}
Bretherton, F.P.
\newblock The motion of rigid particles in a shear flow at low Reynolds number.
\newblock {\em Journal of Fluid Mechanics} {\bf 1962}, {\em 14},~284--304.

\bibitem[Saffman(1965)]{Saffman1965}
Saffman, P.
\newblock The lift on a small sphere in a slow shear flow.
\newblock {\em Journal of Fluid Mechanics} {\bf 1965}, {\em 22},~385--400.

\bibitem[Kim and Yoo(2009)]{KimYoo2009}
Kim, Y.W.; Yoo, J.Y.
\newblock Axisymmetric flow focusing of particles in a single microchannel.
\newblock {\em Lab on a Chip} {\bf 2009}, {\em 9},~1043--1045.

\bibitem[Yuan \em{et~al.}(2016)Yuan, Pan, Zhang, Yan, Zhao, Alici, and
  Li]{YuanLi2016}
Yuan, D.; Pan, C.; Zhang, J.; Yan, S.; Zhao, Q.; Alici, G.; Li, W.
\newblock Tunable particle focusing in a straight channel with symmetric
  semicircle obstacle arrays using electrophoresis-modified inertial effects.
\newblock {\em Micromachines} {\bf 2016}, {\em 7},~195.

\bibitem[Choudhary \em{et~al.}(2019)Choudhary, Renganathan, and
  Pushpavanam]{ChoudharyPushpavanam2019}
Choudhary, A.; Renganathan, T.; Pushpavanam, S.
\newblock Inertial migration of an electrophoretic rigid sphere in a
  two-dimensional Poiseuille flow.
\newblock {\em Journal of Fluid Mechanics} {\bf 2019}, {\em 874},~856--890.

\bibitem[Bazaz \em{et~al.}(2020)Bazaz, Mashhadian, Ehsani, Saha, Kr{\"u}ger,
  and Warkiani]{BazazWarkiani2020}
Bazaz, S.R.; Mashhadian, A.; Ehsani, A.; Saha, S.C.; Kr{\"u}ger, T.; Warkiani,
  M.E.
\newblock Computational inertial microfluidics: a review.
\newblock {\em Lab on a Chip} {\bf 2020}, {\em 20},~1023--1048.

\bibitem[Shin and Sung(2011)]{ShinSung2011}
Shin, S.J.; Sung, H.J.
\newblock Inertial migration of an elastic capsule in a Poiseuille flow.
\newblock {\em Physical Review E} {\bf 2011}, {\em 83},~046321.

\bibitem[Asmolov \em{et~al.}(2018)Asmolov, Dubov, Nizkaya, Harting, and
  Vinogradova]{AsmolovVinogradova2018}
Asmolov, E.S.; Dubov, A.L.; Nizkaya, T.V.; Harting, J.; Vinogradova, O.I.
\newblock Inertial focusing of finite-size particles in microchannels.
\newblock {\em Journal of Fluid Mechanics} {\bf 2018}, {\em 840},~613--630.

\bibitem[Schaaf and Stark(2017)]{SchaafStark2017}
Schaaf, C.; Stark, H.
\newblock Inertial migration and axial control of deformable capsules.
\newblock {\em Soft matter} {\bf 2017}, {\em 13},~3544--3555.

\bibitem[Li \em{et~al.}(2015)Li, McKinley, and Ardekani]{LiArdekani2015}
Li, G.; McKinley, G.H.; Ardekani, A.M.
\newblock Dynamics of particle migration in channel flow of viscoelastic
  fluids.
\newblock {\em Journal of Fluid Mechanics} {\bf 2015}, {\em 785},~486--505.

\bibitem[Raoufi \em{et~al.}(2019)Raoufi, Mashhadian, Niazmand, Asadnia,
  Razmjou, and Warkiani]{RaoufiWarkiani2019}
Raoufi, M.A.; Mashhadian, A.; Niazmand, H.; Asadnia, M.; Razmjou, A.; Warkiani,
  M.E.
\newblock Experimental and numerical study of elasto-inertial focusing in
  straight channels.
\newblock {\em Biomicrofluidics} {\bf 2019}, {\em 13},~034103.

\bibitem[Prohm \em{et~al.}(2013)Prohm, Tr{\"o}ltzsch, and
  Stark]{ProhmStark2013}
Prohm, C.; Tr{\"o}ltzsch, F.; Stark, H.
\newblock Optimal control of particle separation in inertial microfluidics.
\newblock {\em The European Physical Journal E} {\bf 2013}, {\em 36},~118.

\bibitem[Prohm and Stark(2014)]{ProhmStark2014}
Prohm, C.; Stark, H.
\newblock Feedback control of inertial microfluidics using axial control
  forces.
\newblock {\em Lab on a Chip} {\bf 2014}, {\em 14},~2115--2123.

\bibitem[Ioffe and Tihomirov(1979)]{IoffeTihomirov1979}
Ioffe, A.D.; Tihomirov, V.
\newblock {\em Theory of extremal problems}; North-Holland,  1979.

\bibitem[Asmolov(1999)]{Asmolov1999}
Asmolov, E.S.
\newblock The inertial lift on a spherical particle in a plane Poiseuille flow
  at large channel Reynolds number.
\newblock {\em Journal of Fluid Mechanics} {\bf 1999}, {\em 381},~63--87.

\bibitem[Hood \em{et~al.}(2015)Hood, Lee, and Roper]{HoodRoper2015}
Hood, K.; Lee, S.; Roper, M.
\newblock Inertial migration of a rigid sphere in three-dimensional Poiseuille
  flow.
\newblock {\em Journal of Fluid Mechanics} {\bf 2015}, {\em 765},~452--479.

\bibitem[Bruus(2008)]{Bruus2008}
Bruus, H.
\newblock {\em Theoretical microfluidics}; Vol.~18, Oxford university press
  Oxford,  2008.

\bibitem[Ho and Leal(1974)]{HoLeal1974}
Ho, B.; Leal, L.
\newblock Inertial migration of rigid spheres in two-dimensional unidirectional
  flows.
\newblock {\em Journal of fluid mechanics} {\bf 1974}, {\em 65},~365--400.

\bibitem[Schonberg and Hinch(1989)]{SchonbergHinch1989}
Schonberg, J.A.; Hinch, E.
\newblock Inertial migration of a sphere in Poiseuille flow.
\newblock {\em Journal of Fluid Mechanics} {\bf 1989}, {\em 203},~517--524.

\bibitem[Gossett \em{et~al.}(2012)Gossett, Tse, Dudani, Goda, Woods, Graves,
  and Di~Carlo]{GossettDiCarlo2012Small}
Gossett, D.R.; Tse, H.T.K.; Dudani, J.S.; Goda, K.; Woods, T.A.; Graves, S.W.;
  Di~Carlo, D.
\newblock Inertial manipulation and transfer of microparticles across laminar
  fluid streams.
\newblock {\em Small} {\bf 2012}, {\em 8},~2757--2764.

\bibitem[Chun and Ladd(2006)]{ChunLadd2006}
Chun, B.; Ladd, A.
\newblock Inertial migration of neutrally buoyant particles in a square duct:
  An investigation of multiple equilibrium positions.
\newblock {\em Physics of Fluids} {\bf 2006}, {\em 18},~031704.

\bibitem[D{\"u}nweg and Ladd(2009)]{DuenwegLadd2009}
D{\"u}nweg, B.; Ladd, A.J.
\newblock Lattice Boltzmann simulations of soft matter systems. In {\em
  Advanced Computer Simulation Approaches for Soft Matter Sciences III};
  Springer,  2009; pp. 89--166.

\bibitem[Nocedal and Wright(2006)]{NocedalWright2006Chap18}
Nocedal, J.; Wright, S., Sequential Quadratic Programming.
\newblock In {\em Numerical optimization}; Springer Science \& Business Media,
  2006; chapter~18.

\bibitem[Kim \em{et~al.}(2016)Kim, Lee, Wu, Nam, Di~Carlo, and Lee]{KimLee2016}
Kim, J.A.; Lee, J.; Wu, C.; Nam, S.; Di~Carlo, D.; Lee, W.
\newblock Inertial focusing in non-rectangular cross-section microchannels and
  manipulation of accessible focusing positions.
\newblock {\em Lab on a Chip} {\bf 2016}, {\em 16},~992--1001.

\bibitem[Korohoda and Wilk(2008)]{KorohodaWilk2008}
Korohoda, W.; Wilk, A.
\newblock Cell electrophoresis—a method for cell separation and research into
  cell surface properties.
\newblock {\em Cellular \& Molecular Biology Letters} {\bf 2008}, {\em
  13},~312--326.

\end{thebibliography}



\end{document}